\documentclass[12pt,showpacs,showkeys]{revtex4}

\usepackage{amssymb}
\usepackage{amsmath}
\usepackage{graphicx}

\newcommand{\bq}{\begin{eqnarray}}
\newcommand{\eq}{\end{eqnarray}}
\newcommand{\bqn}{\begin{eqnarray*}}
\newcommand{\eqn}{\end{eqnarray*}}
\newcommand{\rr}{{\mathbf r}}

\begin{document}
%%%%%%%%%%%%%%%%%%%%%%%%%%%%%%%%%%%%%%%%%%%%%%%%%%%%%%%%%%%%%%%%%%%%%%%%%%%%%%%

%%%%%%%%%%%%%%%%%%%%%%%%%%%%%%%%%%%%%%%%%%%%%%%%%%%%%%%%%%%%%%%%%%%%%%%%%%%%%%%

%%%%%%%%%%%%%%%%%%%%%%%%%%%%%%%%%%%%%%%%%%%%%%%%%%%%%%%%%%%%%%%%%%%%%%%%%%%%%%%

\title{Computer simulation study of the closure relations in 
hard sphere fluids}

\author{R. Fantoni\footnote{e-mail: {\rm rfantoni@ts.infn.it}}}
%\email{rfantoni@ts.infn.it}
\affiliation{Dipartimento di Fisica Teorica dell' Universit\`a  
and Istituto Nazionale di Fisica della Materia, Strada Costiera 11, 
34014 Trieste, Italy}
\author{G. Pastore\footnote{e-mail: {\rm pastore@ts.infn.it}}}
%\email{pastore@ts.infn.it} 
\affiliation{Dipartimento di Fisica Teorica dell' Universit\`a  
and INFM DEMOCRITOS National Simulation Center, Strada Costiera 11, 
34014 Trieste, Italy}
\date{\today}

\begin{abstract}
\noindent
We study, using Monte
Carlo simulations, the cavity and the bridge functions of 
various hard sphere fluids: 
one component system,
equimolar additive and non additive binary mixtures. 
In particular,  we numerically check  the 
assumption of local dependency 
of the bridge functions from the indirect correlation functions, on 
which most of the existing integral equation theories hinge.
We find that this condition can be violated
either in the region around the first and second neighbors shell, 
or  inside  the hard core,
for the systems here considered. 
The violations manifest themselves clearly in the so called 
Duh-Haymet plots of the bridge functions versus the indirect 
correlation functions and become amplified   as the coupling 
of the system increases. 

\end{abstract}



\maketitle
%%%%%%%%%%%%%%%%%%%%%%%%%%%%%%%%%%%%%%%%%%%%%%%%%%%%%%%%%%%%%%%%%%%%%%%%%%%%%%%
\newpage
\section{Introduction}

A central problem in the theory of the static structure of classical 
liquids is to find a simple and efficient way to obtain   the pair 
correlation functions from the inter-particle forces in pairwise 
interacting fluids. Exact statistical mechanics 
\cite{Hansen,Caillol02} 
allows to write the 
formal solution of such problem as the coupled set of equations:
\begin{equation}
1 + h_{ij}(r) = \exp[-\beta \phi_{ij}(r) + h_{ij}(r) - c_{ij}(r) + 
B_{ij}(r)] \label{gr}
\end{equation}
and
\begin{equation}
h_{ij}(r) = c_{ij}(r) + \sum_{l} \rho_{l} \int d\rr^\prime 
c_{il}(r^\prime)   h_{lj}(|\rr-\rr^\prime |)   \label{OZ}
\end{equation}
where $h_{ij}(r)$  and $c_{ij}(r)$ are the (total) and indirect 
correlation functions for atomic pairs of species $i$ and $j$, 
$\rho_{l}$ is the number density of the $l$-th component and $\beta = 
1/kT$. The functions $B_{ij}(r)$, named bridge functions after their 
diagrammatic characterization \cite{Hansen} are {\em functionals} of
the total correlation functions, i.e. their value at  distance $r$
depends on the values of all the correlation functions at all
distances.  

The basic difficulty with equations (\ref{gr}) and (\ref{OZ}) is that
we do not have an explicit and computationally efficient relation 
between 
$B_{ij}(r)$ and the correlation functions, 
so we have to  resort to approximations. 
The results of the last three decades of research have shown that it 
is possible to make progress by approximating the bridge functionals
$B_{ij}(r)$ by {\em functions } of the indirect correlation functions 
$\gamma_{ij}(r) = h_{ij}(r) - c_{ij}(r)$ (approximate closures). 
Once we have an explicit form for $B_{ij}(\gamma_{ij}(r))$, 
the resulting integral
equations (\ref{gr}) and (\ref{OZ}), although approximate, can
provide excellent results for the static structure of
liquids. Moreover, besides the original focus on the structural
properties, in recent years interest has grown toward using
approximate integral equations to obtain thermodynamics and the phase 
diagrams of liquids and liquid mixtures \cite{Caccamo}.

In particular,  Kjellander and Sarman
\cite{Kjellander89} and Lee \cite{Lee92} have derived
an approximate but useful formula for the chemical potential 
of a fluid requiring only the knowledge of the correlation 
functions at the thermodynamic state of interest. Their formula is 
based on two main approximations. The first is the same  assumption from 
which
integral equations are derived, i.e. that the bridge 
functions $B_{ij}(r)$ are local functions of 
the corresponding indirect correlation functions.
The second stronger 
assumption is that the only dependence  of the bridge
functions from the thermodynamic state is through the indirect 
correlation functions. Thus, the functional dependence of
$B_{ij}(\gamma_{ij})$ is the same for all the states.

In this paper we want to investigate via direct numerical
computer simulation the two approximations.

Up to now, numerical studies of the bridge functions and of the
accuracy of the local approximation have been limited to the case of 
one component systems \cite{Restrepo92,Duh95} or 
electrolytic solutions \cite{Duh92}. 
We feel that two-component 
systems deserve more interest for many reasons: i) there are strong 
indications that the approximate universality of the bridge functions 
\cite{Rosenfeld79} is not valid in multicomponent systems, ii) the 
phase diagrams of multicomponent systems are richer and more
interesting than those of pure fluids, and iii) it turns out that
modeling the bridge functions for multicomponent systems is much more 
difficult than for pure systems.

We have  studied, through Monte Carlo simulation, the
bridge functions of a few systems of non-additive hard spheres (NAHS) 
mixtures, including the limiting cases of additive (AHS) mixtures and
one component system.  
In particular we are interested in a direct check of the local 
hypothesis for the functional relations between bridge and 
correlation functions in binary mixtures. To this aim we use the so 
called Duh-Haymet plots 
\cite{Duh92}. These are 
plots of the partial bridge functions $B_{ij}$ as a function of the
partial indirect correlation functions $\gamma_{ij}$.

The paper is organized as follows.  In section \ref{sec:cavity} we
summarize the equations we used to evaluate the cavity correlation
functions from which the bridge functions can be easily obtained and
we provide the relevant technical details of the numerical
calculations. In section \ref{sec:res} we present and discuss our
numerical results.

%%%%%%%%%%%%%%%%%%%%%%%%%%%%%%%%%%%%%%%%%%%%%%%%%%%%%%%%%%%%%%%%%%%%%%%%%%%%%%%

\section{Calculation of the cavity and bridge functions}
\label{sec:cavity}
\subsection{Theory}

The binary NAHS system is a fluid made of hard spheres 
of two species. One specie, here named 1, with
diameter $R_{11}$ and number density $\rho_1$ and another
specie (2) with diameter $R_{22}$ and number density $\rho_2$,
with a pair interaction potential
that can be written as follows 
\bq \label{pp}
\phi_{ab}(r)=\left\{\begin{array}{ll}
\infty & r<R_{ab}\\
0      & r>R_{ab}
\end{array}\right.~,
\eq
where $R_{12}=(R_{11}+R_{22})/2+\alpha$, with $\alpha$ being the non
additivity parameter. We will also study various special cases as the
one component system,  and the binary mixture of additive
hard spheres (AHS) $\alpha=0$.
We can rewrite Eq. (\ref{gr}) to obtain  the partial bridge functions 
\bq \label{loc:B}
B_{ab}(r)=\ln y_{ab}(r)-\gamma_{ab}(r)~,
\eq 
where $y_{ab}(r)$ are the partial cavity functions
\bq
y_{ab}(r)= g_{ab}(r)\exp[\beta\phi_{ab}(r)]~,
\eq
here $g_{ij}(r) = 1 + h_{ij}(r)$ are the partial radial distribution
functions. Notice that both the cavity functions and the indirect
correlation functions are everywhere continuous then also the bridge
is so. 

In the region outside the hard cores, in a hard sphere (HS) system,
the cavity correlation functions coincide with the pair distribution
functions $g_{ij}(r)$.
In order to determine the relationship between the partial bridge
functions and the partial indirect correlation functions within the
hard cores, we need to calculate the partial cavity functions. 
There are two distinct methods for calculating them \cite{Restrepo92}:
the one which uses Henderson's 
equation \cite{Henderson83} and the direct simulation method of Torrie
and Patey \cite{Torrie77}. We decided to use the first method which 
is accurate at small $r$. 

For a binary mixture the like cavity functions can be obtained from
the following canonical average 
\bq \nonumber
y_{aa}(r_{1_a 2_a})&=&\frac{Vz_a}{N_a}
\bar{y}_{aa}(r_{1_a 2_a})\\ \label{loc:y11}
&=&\frac{Vz_a}{N_a}\left\langle
\exp\left\{-\beta\left[\sum_{i_a>2}^{N_a+1}\phi_{aa}
(r_{1_a i_a})+\sum_{i_b=1}^{N_b}\phi_{ab}
(r_{1_a i_b}) \right]\right\}\right\rangle_{N_1,N_2,V,T}~,
\eq
where $a,b=1,2$ with $b\neq a$, $r_{i_a
j_b}$ is the distance between particle $i$ of specie $a$ and
particle $j$ of specie $b$,
$z_a=\exp(\beta\mu_a)/\Lambda^3$ is the activity of specie
$a$, $\mu_a$ its chemical potential, and $\Lambda$ the de Broglie
thermal wavelength, $V$ is the volume, $N_a$ the
number of particles of specie $a$, so that the prefactor
$Vz_a/N_a = \exp(\beta\mu^{exc}_a)$ where
$\mu^{exc}_a$ is the excess chemical potential of specie
$a$. The notation $\langle\ldots\rangle_{N_1,N_2,V,T}$ indicates
the canonical average at fixed number of particles, volume and
temperature.

So to calculate $\bar{y}_{aa}(r)$ we need to introduce in
the system of $N_a+N_b$ particles labeled
$1_b,\ldots,N_b,2_a,\ldots,(N+1)_a$ 
a test particle $1_a$ placed a distance $r$ from particle
$2_a$ and calculate, at each Monte Carlo step, the interaction of this
particle with all the particles of the system except particle $2_a$. 

We immediately realize that when $r=0$ we must have
\bq \label{loc:ycontact}
\bar{y}_{aa}(0)=1~,
\eq
since the configurations where particle $2_a$ overlaps with other
particles of the system are forbidden. 
Moreover, by taking into account that $y_{ab}(r) = g_{ab}(r)$  for 
$r > R_{ab}$ and from the asymptotic value of the partial pair
distribution functions follows that 
\bq \label{loc:y11asymptotic}
\lim_{r\to\infty}\bar{y}_{aa}(r)=e^{-\beta\mu^{exc}_a}~.
\eq

The unlike cavity functions can be obtained from the following
canonical average  
\bq \nonumber
y_{12}(r_{1_1 1_2})&=&\frac{Vz_1}{N_1}
\bar{y}_{12}(r_{1_1 1_2})\\ \label{loc:y12}
&=&\frac{Vz_1}{N_1}\left\langle
\exp\left\{-\beta\left[\sum_{i_2>1}^{N_2}\phi_{12}
(r_{1_1 i_2})+\sum_{i_1>1}^{N_1+1}\phi_{11}
(r_{1_1 i_1}) \right]\right\}\right\rangle_{N_1,N_2,V,T}~,
\eq

So to calculate $\bar{y}_{12}(r)$ we need to introduce in the system
of $N_1+N_2$ particles labeled $1_2,\ldots,N_2,2_1,\ldots,(N+1)_1$ 
a test particle $1_1$ placed a distance $r$ from particle
$1_2$ and calculate, at each Monte Carlo step, the interaction of this
particle with all the particles of the system except particle $1_2$.

Now there is no simple argument to guess the contact value of
$\bar{y}_{12}$. All we can say is that we must have
$\bar{y}_{12}(0)\le 1$. At large $r$ we still have 
\bq \label{loc:y12asymptotic}
\lim_{r\to\infty}\bar{y}_{12}(r)=e^{-\beta\mu^{exc}_1}~.
\eq

%%%%%%%%%%%%%%%%%%%%%%%%%%%%%%%%%%%%%%%%%%%%%%%%%%%%%%%%%%%%%%%%%%%%%%%%%%%%%%%

\subsection{Numerical implementation}

Monte Carlo simulations were performed with a standard NVT
Metropolis algorithm \cite{AT} using $N=4000$ particles.  
Linked lists  \cite{AT} have been used to reduce the computational
cost. To calculate the partial pair distribution functions we
generally used $5.2\times 10^8$ Monte Carlo steps, where one step
corresponds to the attempt to move a single randomly chosen particle, 
and incremented
the histograms once every $20\times 4000$ steps. To calculate the
partial cavity functions we used $1.6\times 10^9$ Monte Carlo steps
and incremented the histograms once every $2\times 4000$ steps. The 
acceptance ratio was adjusted to values between 10\% and 40\%.

The Monte Carlo simulation returned the $g_{ab}(r)$ over a range not
less than $8.125R_{11} $ for the densest system. In all the studied
cases, the pair distribution functions attained their asymptotic value
well inside the maximum distance they were evaluated. Thus, it has
been possible to obtain accurate Fourier transforms of the total
correlation functions [$\hat h_{ab}(k)$] (it was necessary to cure the
cusps at contact in the partial pair distribution functions by
adding to them $H(R_{ab}-r)g_{ab}(R_{ab})$, $H$ being the Heaviside
step function, before taking the Fourier transform and removing 
its analytical Fourier transform
afterwards). To obtain the partial indirect correlation functions we
first calculated the partial direct correlation functions
[$\hat c_{ab}(k)$] using the Fourier transform of the Ornstein-Zernike 
equation (\ref{OZ}) and then we got the Fourier transform of the 
indirect correlation 
functions $\hat \gamma_{ij}(k)=\hat h_{ij}(k)-\hat c_{ij}(k)$ 
which is the transform of a continuous function in real space and 
then is safe to transform back numerically to obtain $\gamma_{ab}(r)$.

%%%%%%%%%%%%%%%%%%%%%%%%%%%%%%%%%%%%%%%%%%%%%%%%%%%%%%%%%%%%%%%%%%%%%%%%%%%%%%%

\section{Numerical Results}
\label{sec:res}

We carried on simulations on the following systems: (A), one component
HS; (B), equimolar binary mixture of AHS; (C), equimolar binary
mixture of NAHS with equal like diameters and negative non additivity;
(D), equimolar binary mixture of NAHS with equal like diameters and
positive non additivity; and (E), equimolar binary mixture of NAHS
with different like diameters. In all these cases we have drawn the 
corresponding Duh-Haymet plots, i.e. we plot, for each distance, the pairs 
$(B_{ij}(r),\gamma_{ij}(r) )$. 

When we are outside the hard core the partial bridge functions
(\ref{loc:B}) reduces to 
\bq
B_{ab}(r)=\ln g_{ab}(r)-\gamma_{ab}(r)~.
\eq
and we can obtain the bridge functions directly from the pair 
correlation functions solving the OZ equation
(\ref{OZ}) to get the partial indirect correlation functions
$\gamma_{ab}$.

To realize the Duh-Haymet plots when we are within the hard core
regions, we first calculated the cavity functions $\bar{y}_{ab}$ as
explained in section \ref{sec:cavity} and then the bridge functions 
(up to  an additive constant, the excess chemical potential
$\beta\mu_a^{exc}~$) from their definition (\ref{loc:B}). 
Estimating the excess chemical potential from the long range behavior
of the cavity functions [see equations (\ref{loc:y11asymptotic}) and
(\ref{loc:y12asymptotic})] we where able to find the full bridge
functions. 
Since the cavity functions in proximity of $R_{ab}$ becomes very
small, they are subject to statistical errors. In order to obtain
smooth Duh-Haymet plots we needed to smooth the cavity functions
obtained from the simulation. We did this by constructing the cubic
smoothing spline which has as small a second derivative as possible.

\subsection{One component HS}
\label{subsec:ochs}

We carried out two simulations at $\rho_1\simeq 0.650$
(with a packing fraction of $\eta=\pi\rho_1R_{11}^3/6=0.340$) and
$\rho_1\simeq 0.925$ ($\eta= 0.484$), the former corresponding to an
intermediate density case and the latter to a liquid close to the
freezing point. In our simulations we use $R_{11}$ as unit of length. 

Inside the hard core, the bridge and the indirect correlation
functions are monotonic and, for the cases here considered, there are
no non-localities in the Duh-Haymet plots inside the core. Thus, to
search for non-localities it is enough to analyze results in the
external region. The resulting curves in the $(B,\gamma)$ plane
corresponding to points outside the hard core region are shown in
Fig. \ref{boc_ochs}. On the left the intermediate  density case and
on the right the high density one. We see that, as the density
increases, the non-locality becomes more accentuated. Of course, the
quality of a local approximation does depend on the choice of the
correlation functions used as independent variable: 
plotting the bridge function as function of the direct correlation
function we observed the opposite behavior.

In order to compare the computer simulation results with the local
approximate $B(\gamma)$ relations used in the integral equations,  
we have plotted the curves corresponding to different closures:
the hyper-netted chain (HNC) \cite{Hansen}:
\bq
B(\gamma) = 0~,
\eq
the Percus Yevick (PY) \cite{Hansen}; 
\bq
B(\gamma) = log(1+\gamma) - \gamma~,
\eq
the  Martynov Sarkisov (MS) \cite{Martynov83} and its generalization
by Ballone, Pastore, Galli, and Gazzillo (BPGG) \cite{Ballone86}: 
\bq
B(\gamma) = (1+\alpha \gamma)^{1/\alpha} - \gamma -1~,
\eq
(MS corresponds to $\alpha=2$, in the BPGG generalization $\alpha$
could be used as state dependent parameter to enforce thermodynamic
consistence, here a fixed  value of $15/8$ has been used as suggested
in \cite{Ballone86}), and the modified Verlet (MV) \cite{Verlet80}:
\bq
B(\gamma) = \frac{-\gamma^2}{2 \left[ 1 + 0.8 \gamma \right] }~.
\eq
We can see that the best closures (MS, BPGG and MV), although not passing 
through the simulation curve, tend to follow its slope and curvature. 
When looking at Fig. \ref{boc_ochs} one should also bear in mind that the 
values of the bridge function outside the hard core are not the most 
relevant for the quality of the structural and thermodynamic results of the 
closures.

\subsection{Equimolar binary mixture of AHS}
\label{subsec:ahs}

We carried out a simulation at $\rho_1=\rho_2\simeq 0.589$
[$\eta=\pi(\rho_1R_{11}^3+\rho_2R_{22}^3)/6= 0.375$] and
$\rho_1=0.5$. We chose $R_{11}=1$, $R_{12}=0.8$, and $R_{22}=0.6$. 

The results outside the hard core region are shown in the insets of 
the plots of Fig. \ref{bvg_ahs-ed}. There are non-localities in a  
neighborhood of the origin which corresponds to the large $r$ region.  
These are more evident in the high density case.

The most interesting feature shown in the figure is the difference 
between the curves at the two different densities. If the hypothesis
of closures defined by a  
unique function $B(\gamma)$ would be exact data for different densities 
should collapse into a unique curve in these plots. 
The data shown in Figs. \ref{boc_ochs} and \ref{bvg_ahs-ed} indicate 
clearly that this not strictly true. However, at low and intermediate 
densities the quantitative effect of the changing functional form is not 
dramatic. And even at the highest liquid densities, the success of
closures such as MV, MS or BPGG can be probably explained in term of a
higher sensitivity of the theory to localized (near the contact)
features of the bridge functions more than to the behavior over the
whole range of distances. 

Inside the hard core region the Duh-Haymet plots do not have 
non-localities. In Fig. \ref{bic_ahs-ed} we show the results for the
cavity functions $\bar{y}_{ab}$ for the system at the highest
density. The plot for the unlike functions is more noisy
than the plots for the like functions because $\bar{y}_{12}$ being 
smaller than $\bar{y}_{aa}$ for $a=1,2$ is more subject to
statistical error.

In Fig. \ref{bvg_ahs-ed} we show the full Duh-Haymet plots for the
system at the highest density, from the simulation (dots) and from
integral equation theories (lines).
The plots show how the MV approximation is the best one for
this system. The unlike bridge function starts at $r=0$ close to the
MV approximation, stays close to this approximation as $r$
increases and at some point have a smooth change in behavior and
get closer to the  PY curve. 

Fig. \ref{br_ahs-ed} shows the full bridge functions as a function of
$r$ for the system at the highest density. It is worth of notice the
almost flat region of the unlike bridge near the origin.

\subsection{Equimolar binary mixture of NAHS: $R_{11}=R_{22}$, $\alpha
< 0$}
\label{subsec:nahs-een}

We carried out a simulation at $\rho_1=\rho_2\simeq 0.573$
($\eta=0.6$). We chose $R_{11}=R_{22}=1$ and $R_{12}=0.649$
($\alpha=-0.351$). These radii values would be  suitable for 
a reference system  to  model correlation in  molten NaCl
\cite{Ballone84}.  

The results outside the hard core region are shown in the insets of
the plots of Fig. \ref{bvg_nahs-een}. There are non-localities in the
neighborhood of the origin corresponding to the large $r$ region.

In Fig. \ref{bic_nahs-een} we show the results for the cavity
functions $\bar{y}_{ab}$. 

In Fig. \ref{bvg_nahs-een} we show the full Duh-Haymet plots from the
Monte Carlo simulation (dots) and from the most common integral
equation theories (lines). The approximation which seems to be closer
to the like bridge function is MV: only at big $r$ the bridge
functions is well approximated by PY, MS, BPGG, and MV. The unlike
bridge function starts at $r=0$ close to the PY approximation but as
$r$ increases it has a sudden change in behavior which displaces it
away from all the approximations. Inside the hard core region the
Duh-Haymet plots for the unlike 
functions exhibit significant non-localities in correspondence with
the non monotonic behavior of the unlike cavity function (black dots
in Fig. \ref{bic_nahs-een}).

Fig. \ref{br_nahs-een} shows the full bridge functions as a function
of $r$. The unlike bridge function shows oscillations in a
neighborhood of the origin.

\subsection{Equimolar binary mixture of NAHS: $R_{11}=R_{22}$, $\alpha
> 0$}
\label{subsec:nahs-eep}

We carried out a simulation at $\rho_1=\rho_2\simeq 0.200$
($\eta=0.209$). We chose $R_{11}=R_{22}=1$ and $R_{12}=1.2$
($\alpha=+0.2$). Notice that this system undergoes phase separation
when $\rho=2\rho_1 > 0.42$. 

The results outside the hard core region are shown in the insets of
the plots of Fig. \ref{bvg_nahs-eep}. There are non-localities in a
neighborhood of the origin corresponding to large distances.

Also for this system, inside the hard core region the Duh-Haymet plots
for the unlike functions have non-localities in a neighborhood of
$r=0$. These are smaller in extent than the ones found for system
C. In Fig. \ref{bic_nahs-eep} we show the results for the cavity 
functions $\bar{y}_{ab}$.

In Fig. \ref{bvg_nahs-eep} we show the full Duh-Haymet plots from
the simulation (dots) and from the most common integral equations
(lines). The approximations which seem to be closer to the
like bridge function is MV and BPGG even if there is always a gap
between the approximations and the simulation. The unlike bridge
function starts at $r=0$ far away from all the approximations but as
$r$ increases it has a smooth change in behavior approaching the BPGG
curve. 

Fig. \ref{br_nahs-eep} shows the full bridge functions as a function 
of $r$. Again, the unlike bridge function have an almost flat behavior
in a neighborhood of the origin. 

\subsection{Equimolar binary mixture of NAHS: $R_{11}\ne R_{22}$}
\label{subsec:nahs-edn}

We carried out a simulation at $\rho_1=\rho_2\simeq 0.589$
($\eta=0.375$). We chose $R_{11}=1$ and $R_{12}=R_{22}=0.6$
($\alpha=-0.2$). 

The results outside the hard core region are shown in the insets of
the plots of Fig. \ref{bvg_nahs-edn}. There are non-localities
in a neighborhood of the origin which corresponds to the big $r$
region. 

Inside the hard core region the Duh-Haymet plots have no
non-localities. In Fig. \ref{bic_nahs-edn} we show the results for the
cavity functions $\bar{y}_{ab}$. 

In Fig. \ref{bvg_nahs-edn} we show the full Duh-Haymet plots from
the simulation (dots) and from the most common integral equations
(lines). The approximation which is closer to the 11 bridge
function is the MV. The one that is closer to the 22 bridge function
is the BPGG. The 12 bridge function starts at $r=0$ far away from all
the 5 approximations and as $r$ increases has a sudden change in
behavior and starts following the BPGG approximation.

Fig. \ref{br_nahs-edn} shows the full bridge functions as a function 
of $r$. The unlike bridge function shows again a qualitatively
different behavior near the origin.

%%%%%%%%%%%%%%%%%%%%%%%%%%%%%%%%%%%%%%%%%%%%%%%%%%%%%%%%%%%%%%%%%%%%%%%%%%%%%%%

\section{Conclusions}

\begin{figure}[hbtp]
\begin{center}
\includegraphics[width=8cm]{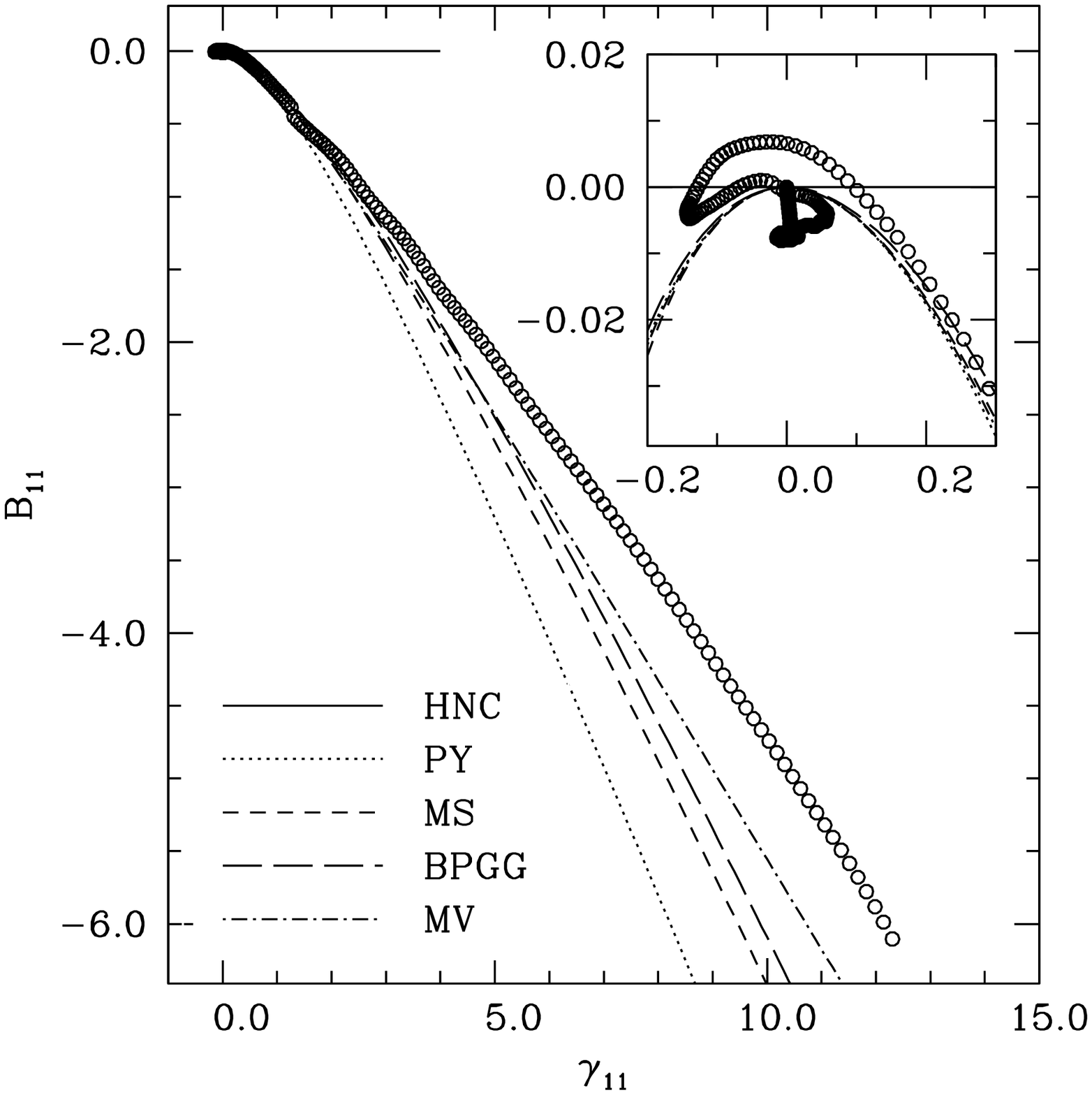}
\includegraphics[width=8cm]{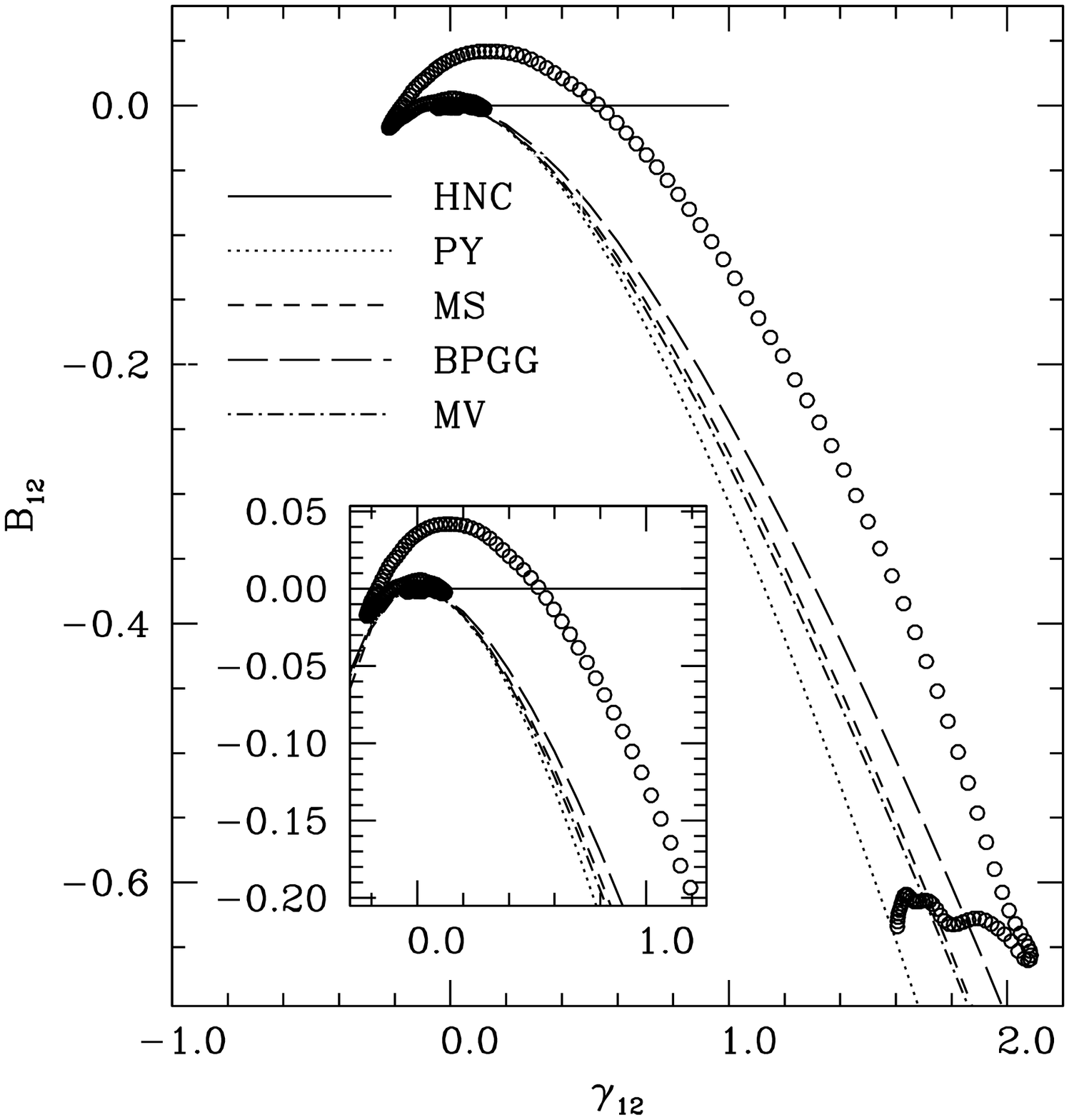}
\end{center}
\caption[]{R. Fantoni and G. Pastore
\label{bvg_nahs-een}
}
\end{figure}

\begin{figure}[hbtp]
\begin{center}
\includegraphics[width=8cm]{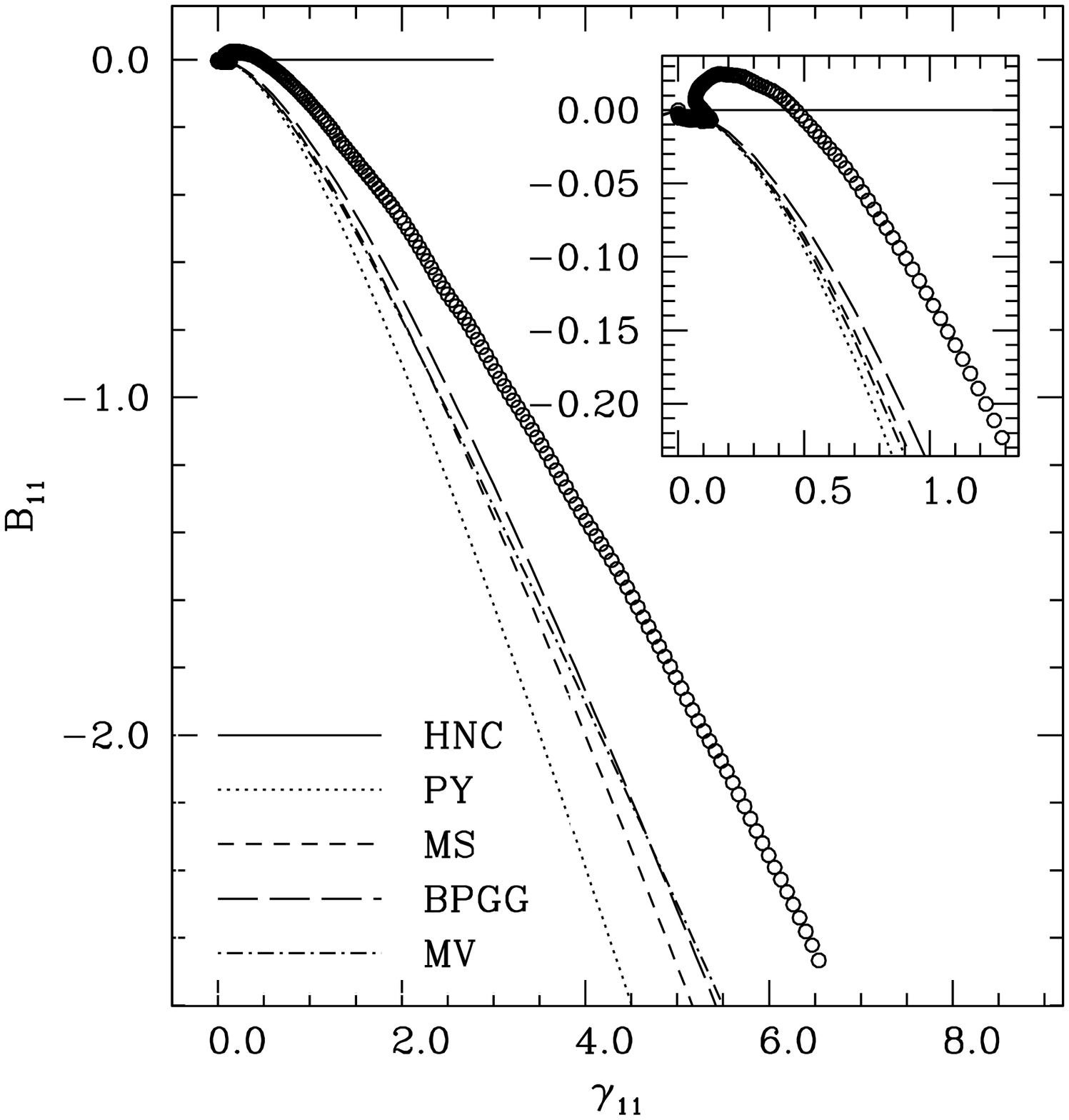}
\includegraphics[width=8cm]{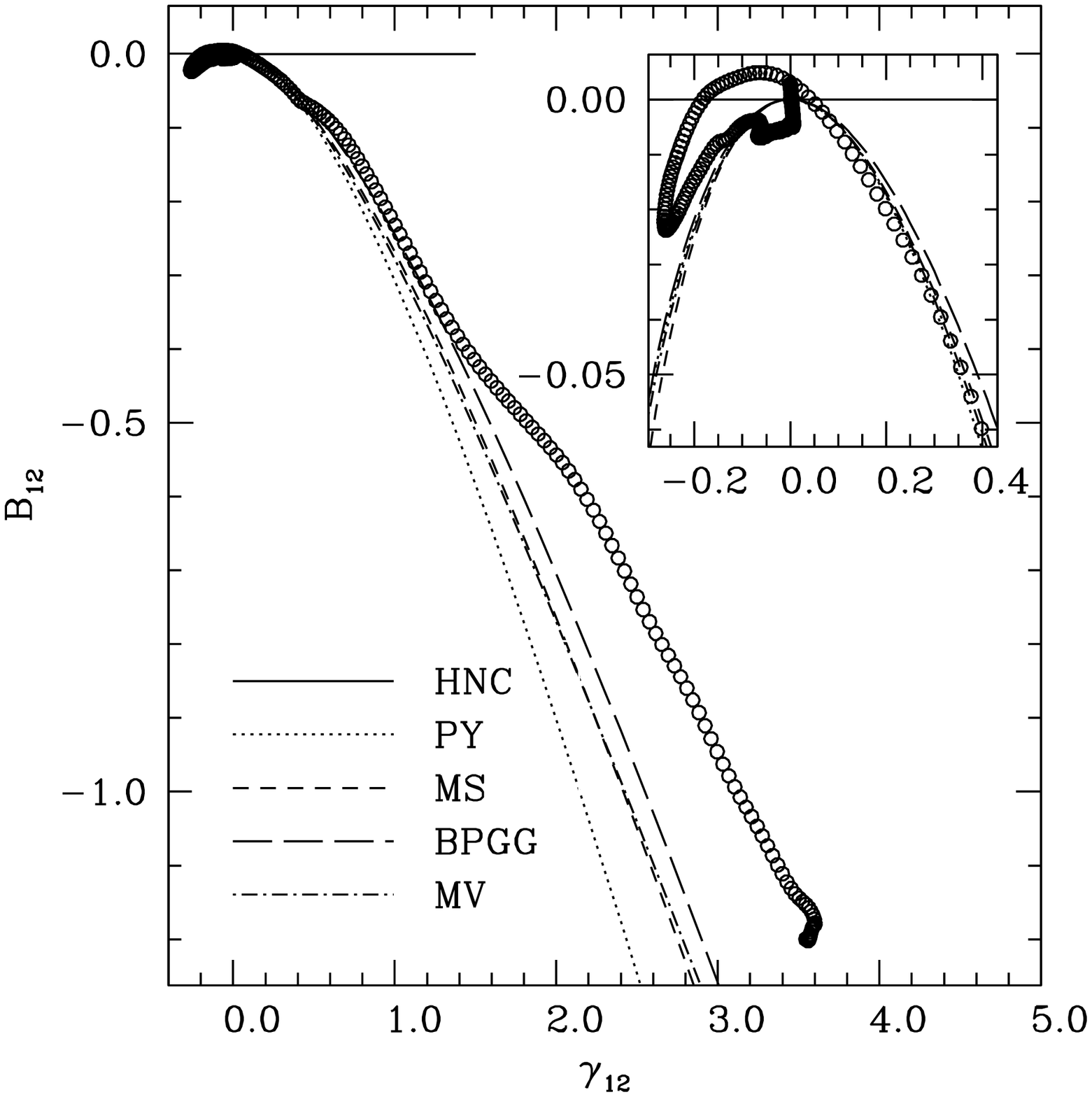}
\end{center}
\caption[]{R. Fantoni and G. Pastore
\label{bvg_nahs-eep}
}
\end{figure}

From our analysis follows that the non-localities in the function
relationship between the bridge functions and the indirect correlation
functions may appear either outside of the hard core regions or
inside of it. While the non-localities outside the hard core appear
both in the like and in the unlike functions, the ones inside the hard
core appear only in the unlike functions (see
Fig. \ref{bvg_nahs-een} and Fig. \ref{bvg_nahs-eep}),
for the systems that we have studied.  Their appearance can be
directly related to the peculiar behavior of the unlike cavity
correlation function inside the hard core.

\begin{figure}[hbtp]
\begin{center}
\includegraphics[width=8cm]{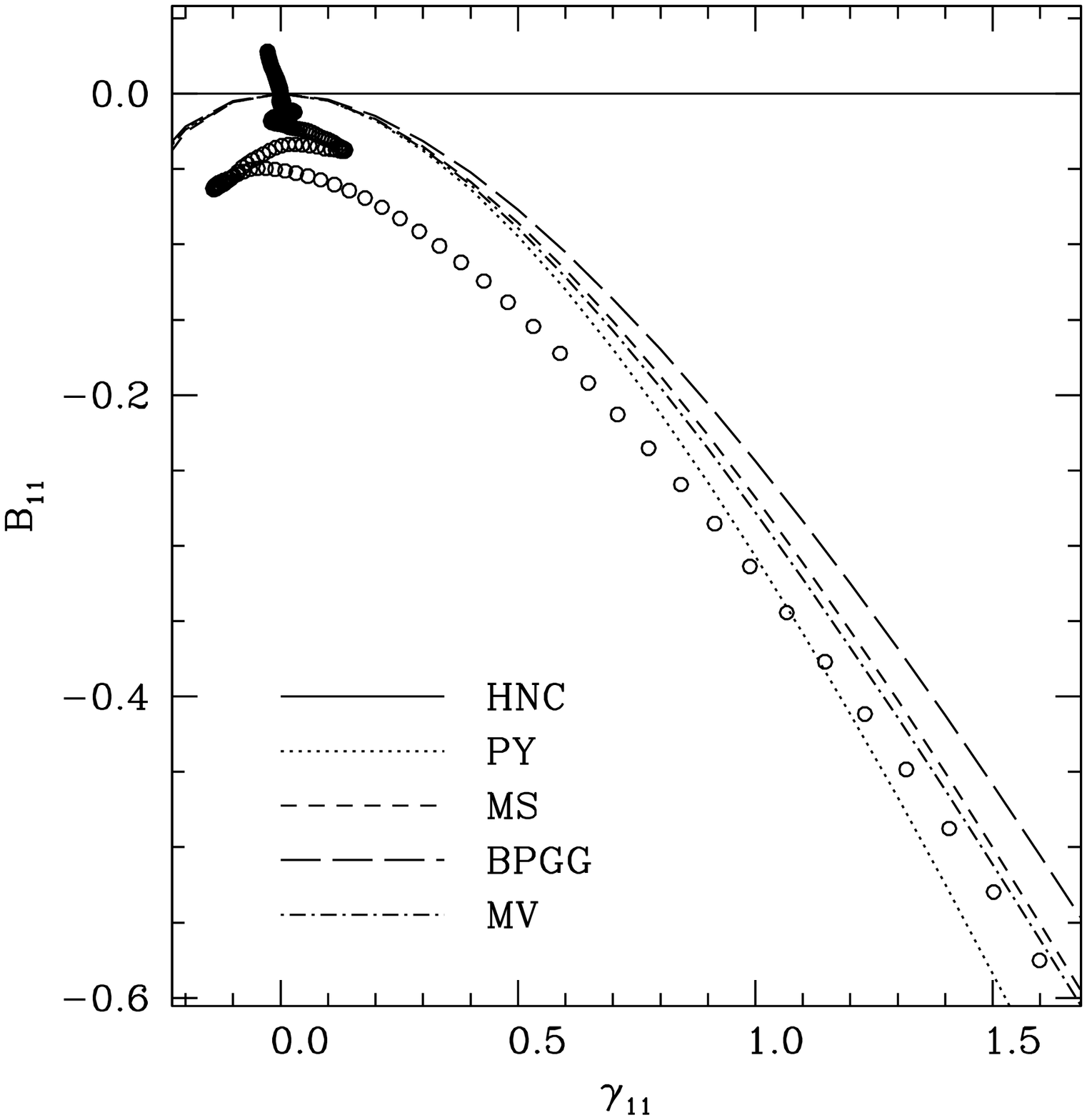}
\includegraphics[width=8cm]{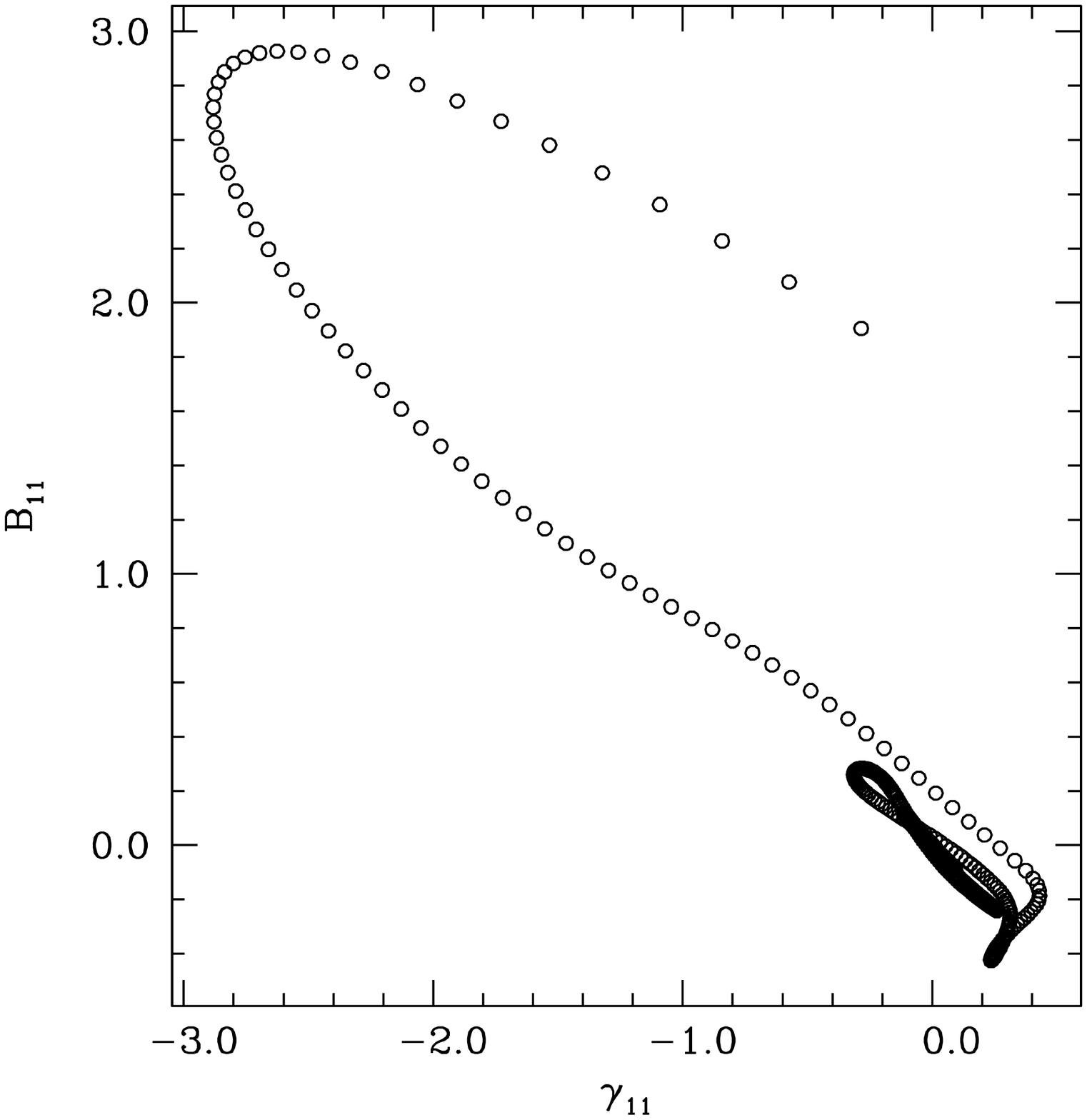}
\end{center}
\caption[]{R. Fantoni and G. Pastore 
\label{boc_ochs}
}
\end{figure}

\begin{figure}[hbtp]
\begin{center}
\includegraphics[width=8cm]{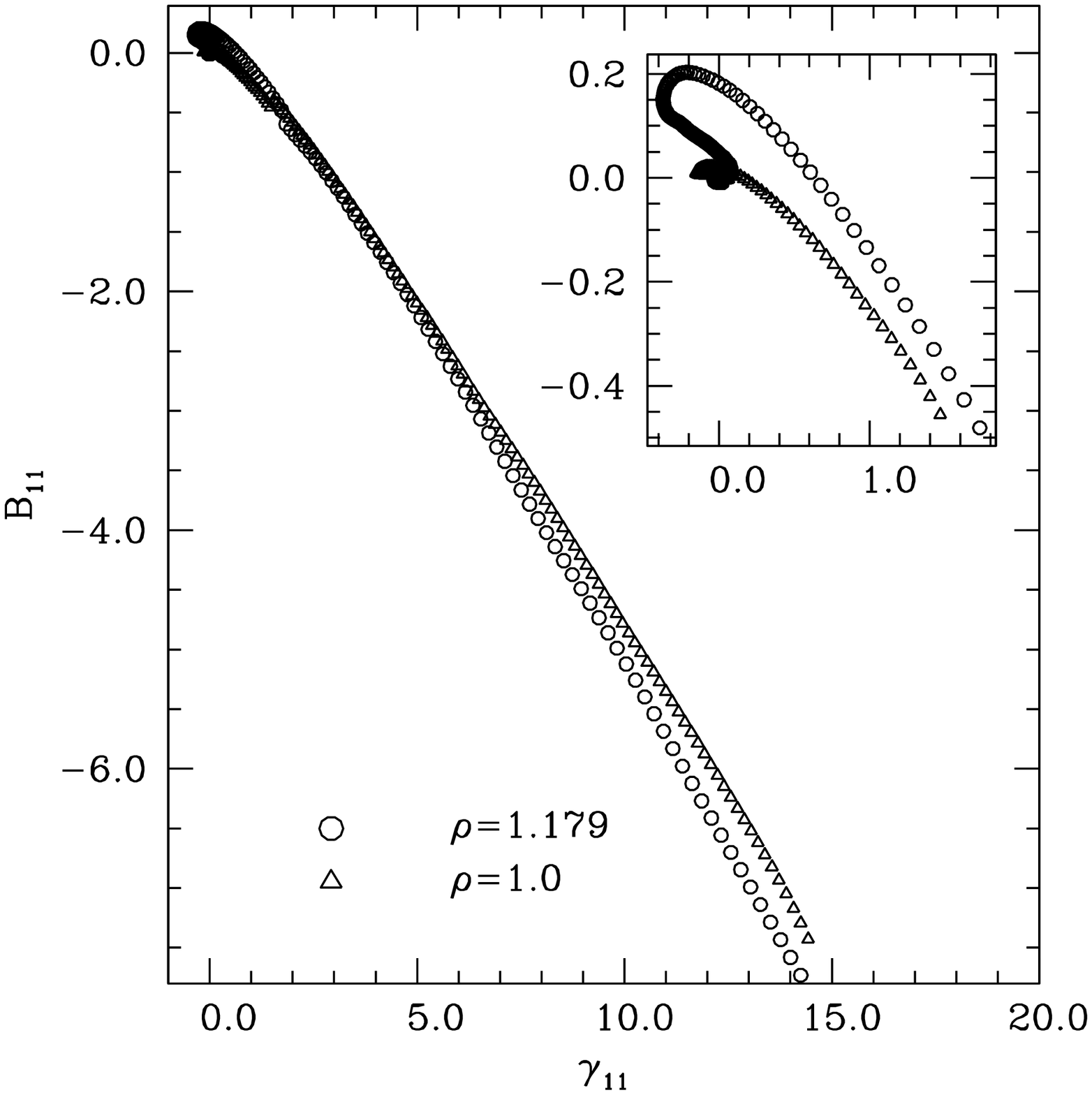}
\includegraphics[width=8cm]{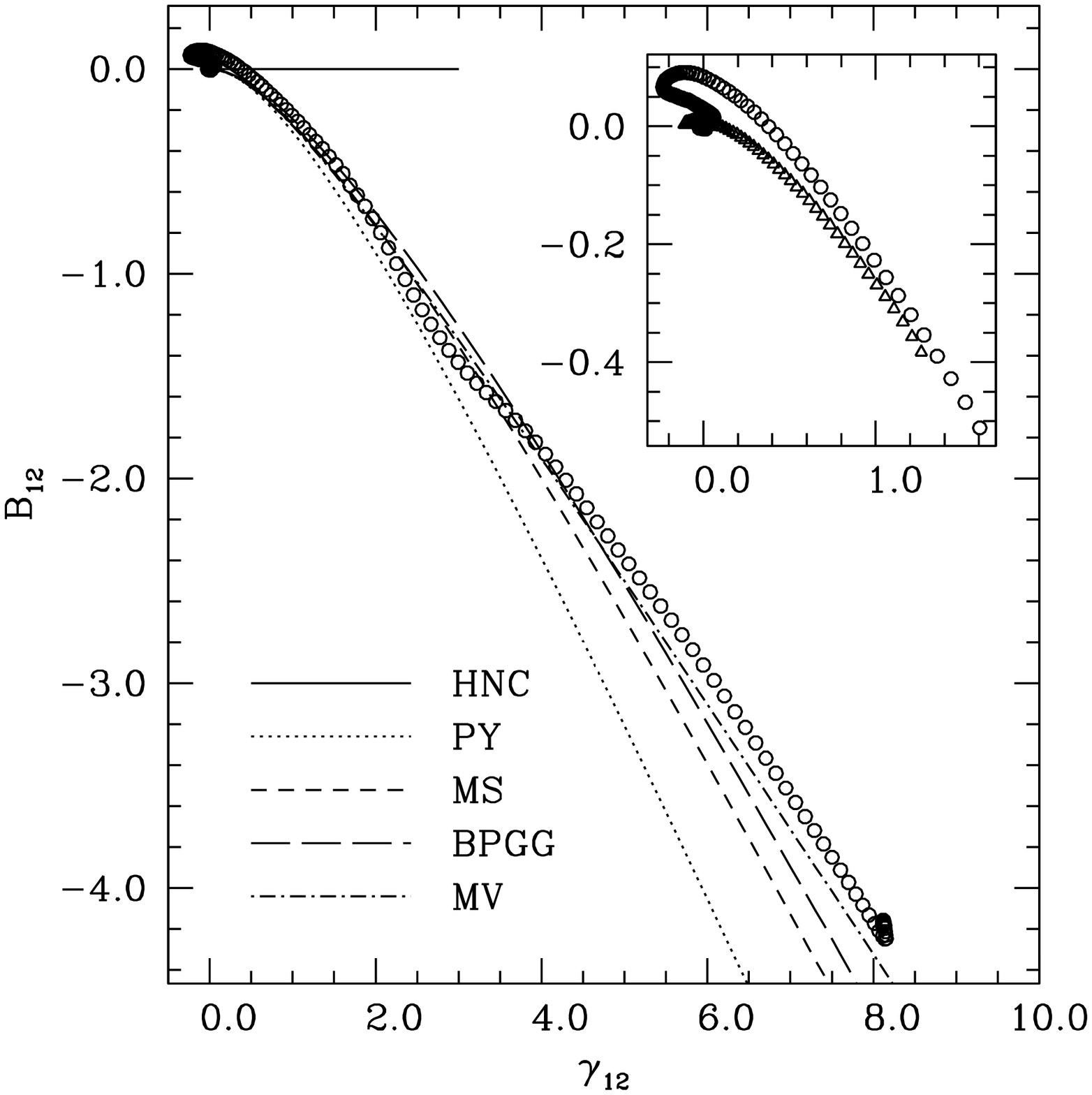}\\
\includegraphics[width=8cm]{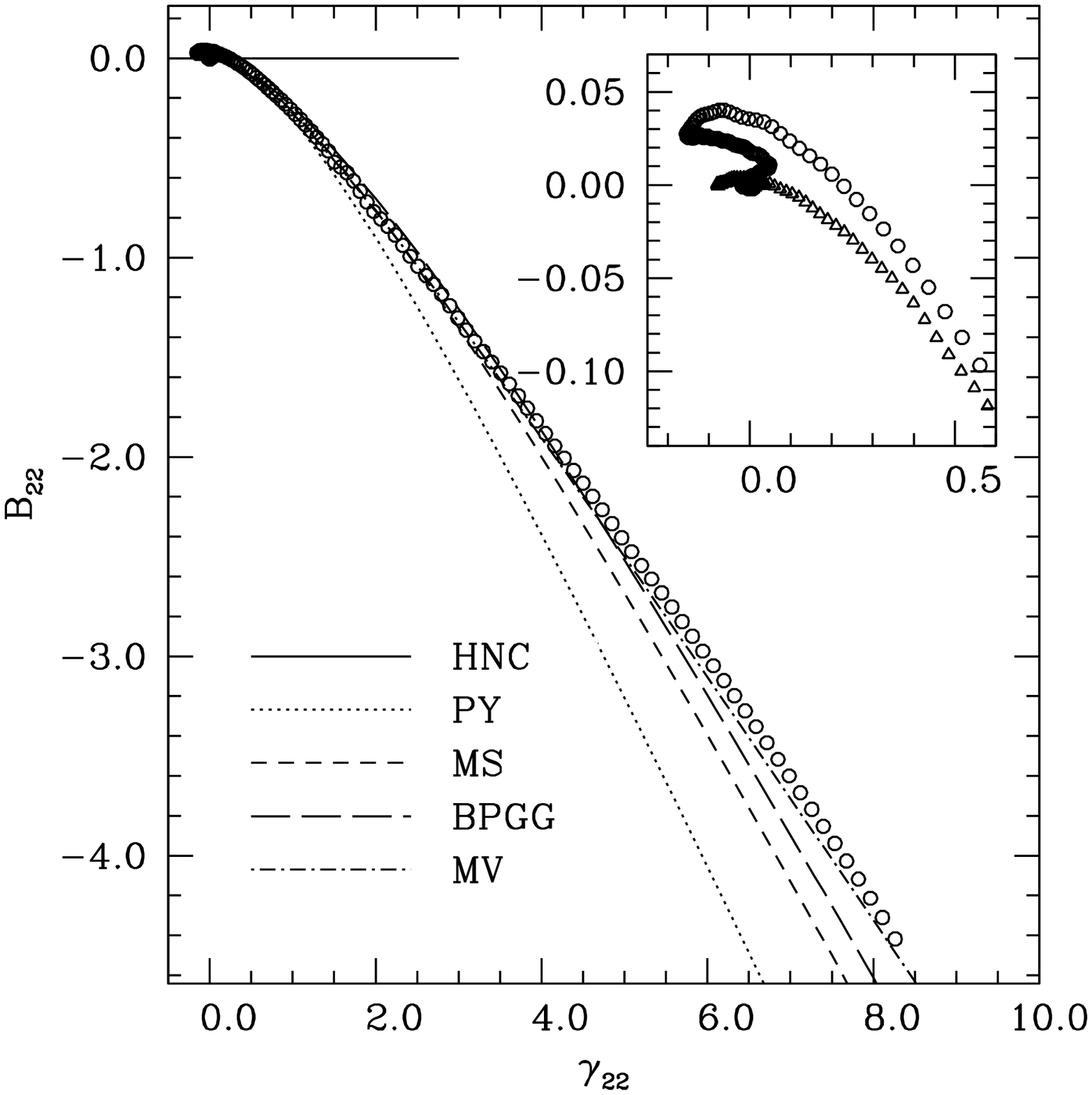}
\end{center}
\caption[]{R. Fantoni and G. Pastore
\label{bvg_ahs-ed}
}
\end{figure}

As is shown by a comparison of the plots of 
Fig. \ref{boc_ochs} and from Fig. \ref{bvg_ahs-ed} the non-localities
become more accentuated as we increase the coupling (the density) of
the system. 
Nonetheless Fig. \ref{bvg_nahs-eep} shows that the
non-localities may appear even in a weakly coupled system (in this
case symmetric NAHS with positive non additivity).
Among the systems studied the one which presents the worst
non-localities is the equimolar symmetric NAHS with negative non
additivity (see Fig. \ref{bvg_nahs-een}) .
For this system the Duh-Haymet plot for the unlike bridge
function is non-local both in the hard core region (in a neighborhood
of $r=0$) and outside of it (at large $r$).

We can conclude that the two hypothesis of a local function
approximation for the bridge functionals of the indirect
correlation functions and the stronger hypothesis of unique functional
form independent on the state, are not strictly supported by the
numerical data. For the one component system, this finding is
consistent with the observed density dependence of the bridge function
reported in \cite{Malijevsky87}. 
We observe clear violations of both the assumptions increasing with
the density.  
This negative statement should be somewhat mitigated by realizing that
the violations of the locality, in the systems studied, are limited to
the small and large distances regions. The latter, corresponding to
the region of the fast vanishing of the bridge functions affect very
little the thermodynamic and structural properties of the systems. The
former are presumably more important for the level of  thermodynamic
consistence of the theory but have small effect on quality of the
structural results. The well known success of closures like MS, BPGG
and MV supports such point of view. 

From comparison with the simulation data in the cases we have studied,
we conclude that the best
approximations  of the true hard sphere bridge functions are provided by the 
MV and BPGG even if, especially in the
unlike bridge functions, there are a wide variety of characteristic behaviors
which are not captured by any of the  most popular integral equation
approximations.
In this respect, we feel that a final comment on the local functional
approximation in the case of multicomponent systems is in
order. Indeed, density functional theory allows to say that the bridge
function $B_{ij}$ should be a functional of all the pair correlation
functions, not only the $(i,j)$ one. Thus, we could have a function
approximation $B_{ij}(\gamma_{11}(r),\gamma_{12}(r),\gamma_{22}(r))$
which would be local in space but not with respect to  the
components. At the best of our knowledge, up to now no attempt has
been done to explore this additional freedom to improve the modeling
of the bridge functions in multicomponent systems.

%%%%%%%%%%%%%%%%%%%%%%%%%%%%%%%%%%%%%%%%%%%%%%%%%%%%%%%%%%%%%%%%%%%%%%%%%%%%%%%

%\begin{acknowledgments}

%\end{acknowledgments}
%%%%%%%%%%%%%%%%%%%%%%%%%%%%%%%%%%%%%%%%%%%%%%%%%%%%%%%%%%%%%%%%%%%%%%%%%%%%%%%

\bibliography{loc2}

%%%%%%%%%%%%%%%%%%%%%%%%%%%%%%%%%%%%%%%%%%%%%%%%%%%%%%%%%%%%%%%%%%%%%%%%%%%%%%%

\begin{figure}[hbtp]
\begin{center}
\includegraphics[width=8cm]{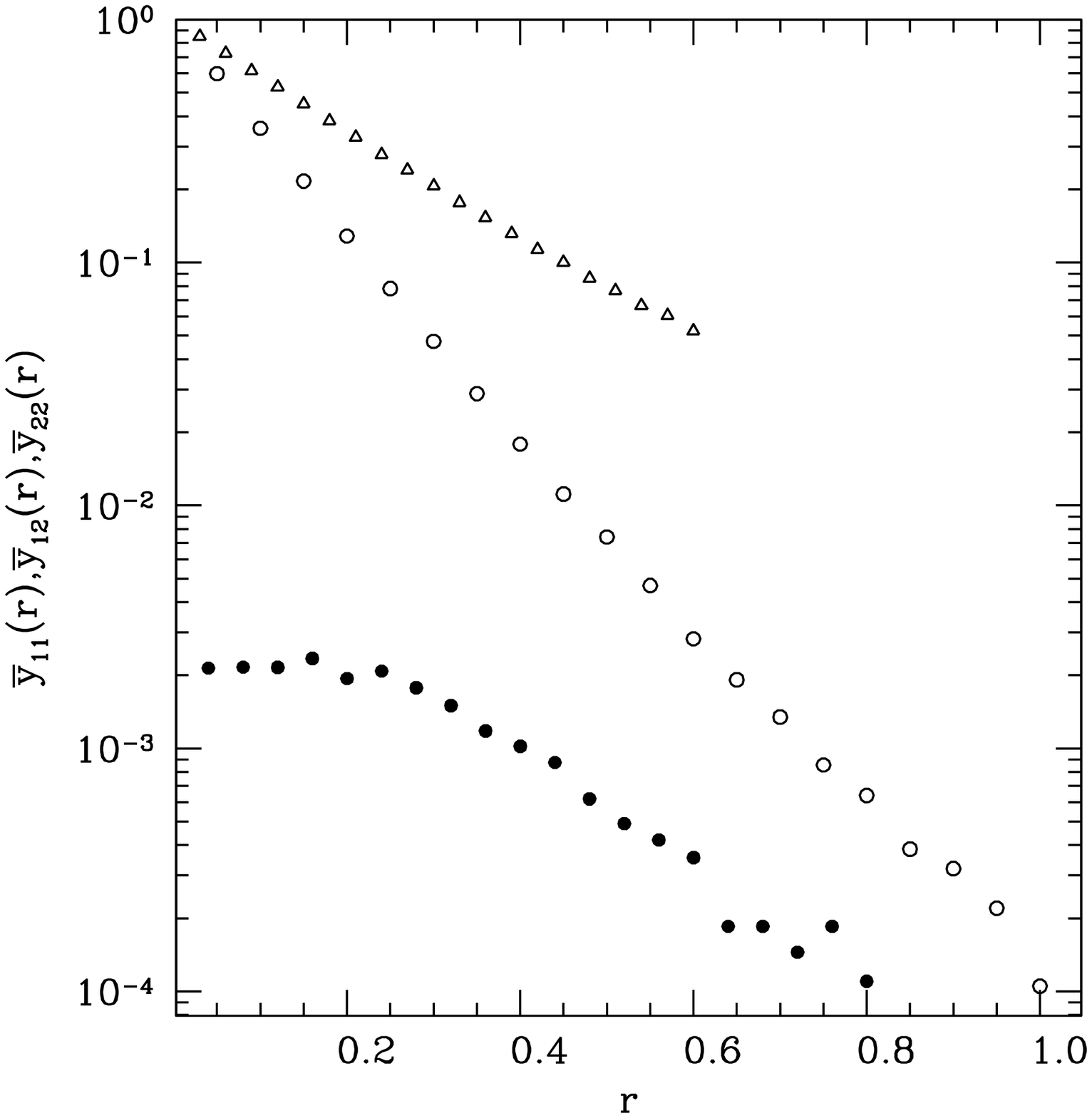}
\end{center}
\caption[]{R. Fantoni and G. Pastore 
\label{bic_ahs-ed}
}
\end{figure}

\begin{figure}[hbtp]
\begin{center}
\includegraphics[width=8cm]{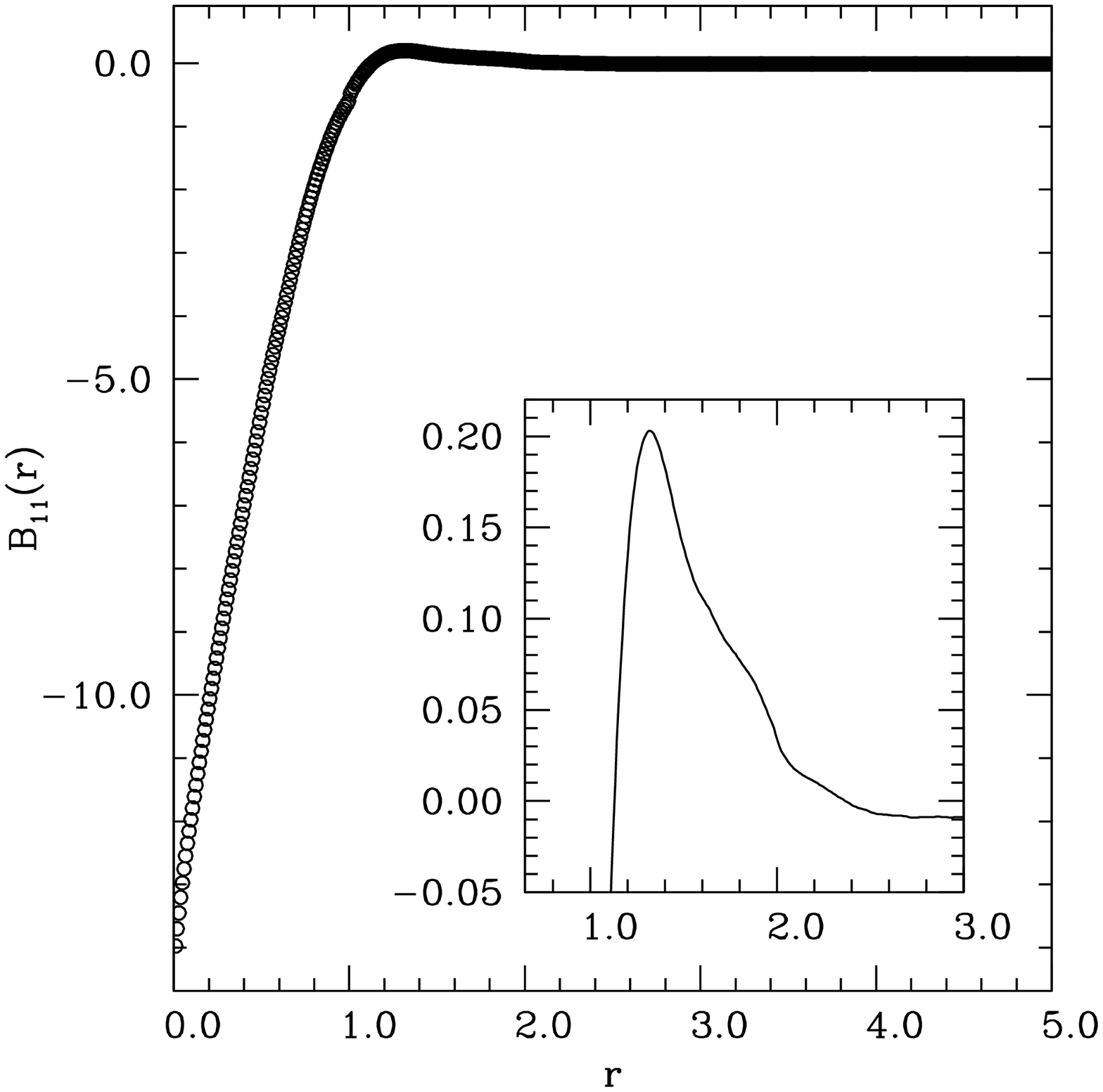}
\includegraphics[width=8cm]{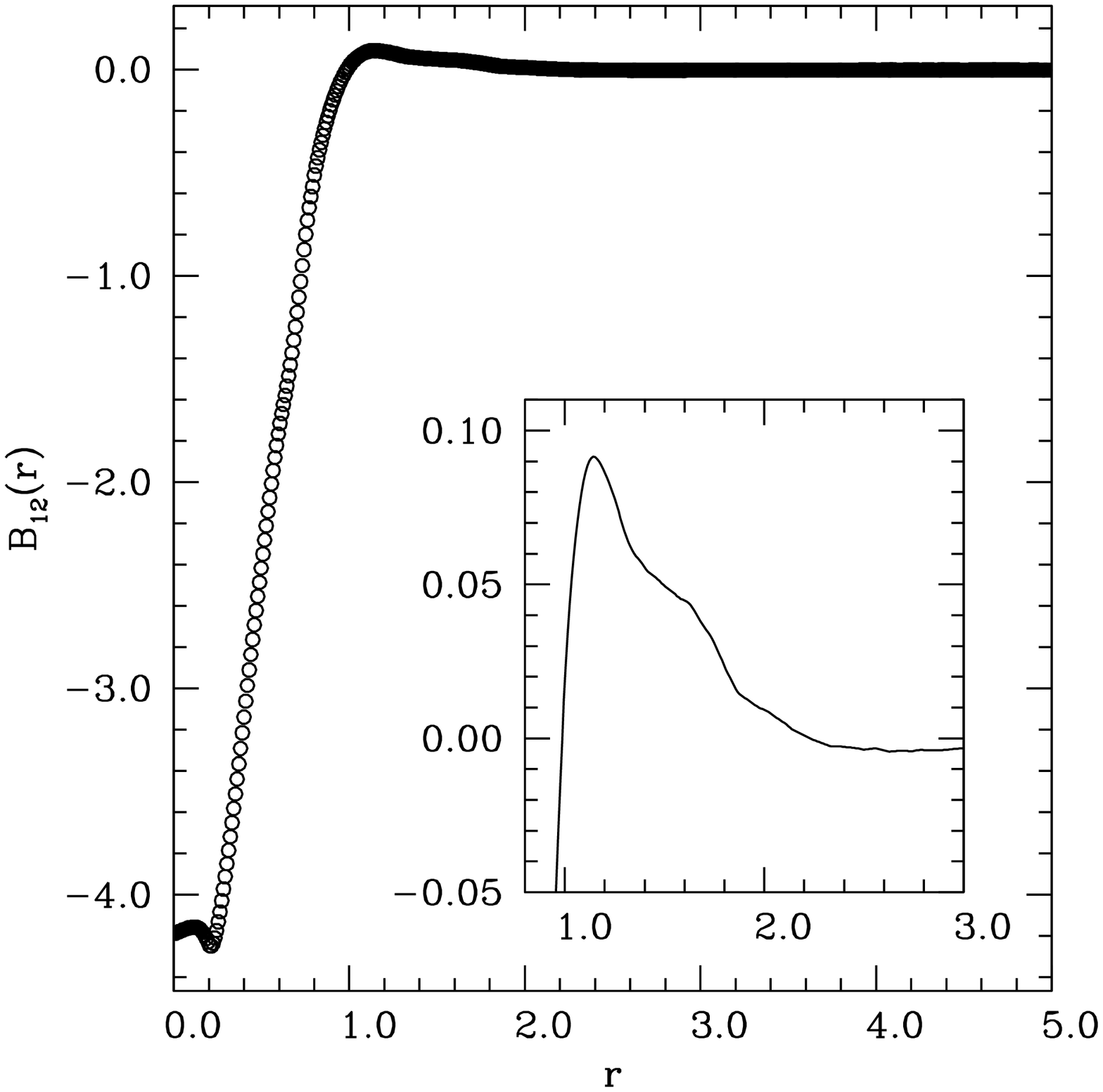}
\includegraphics[width=8cm]{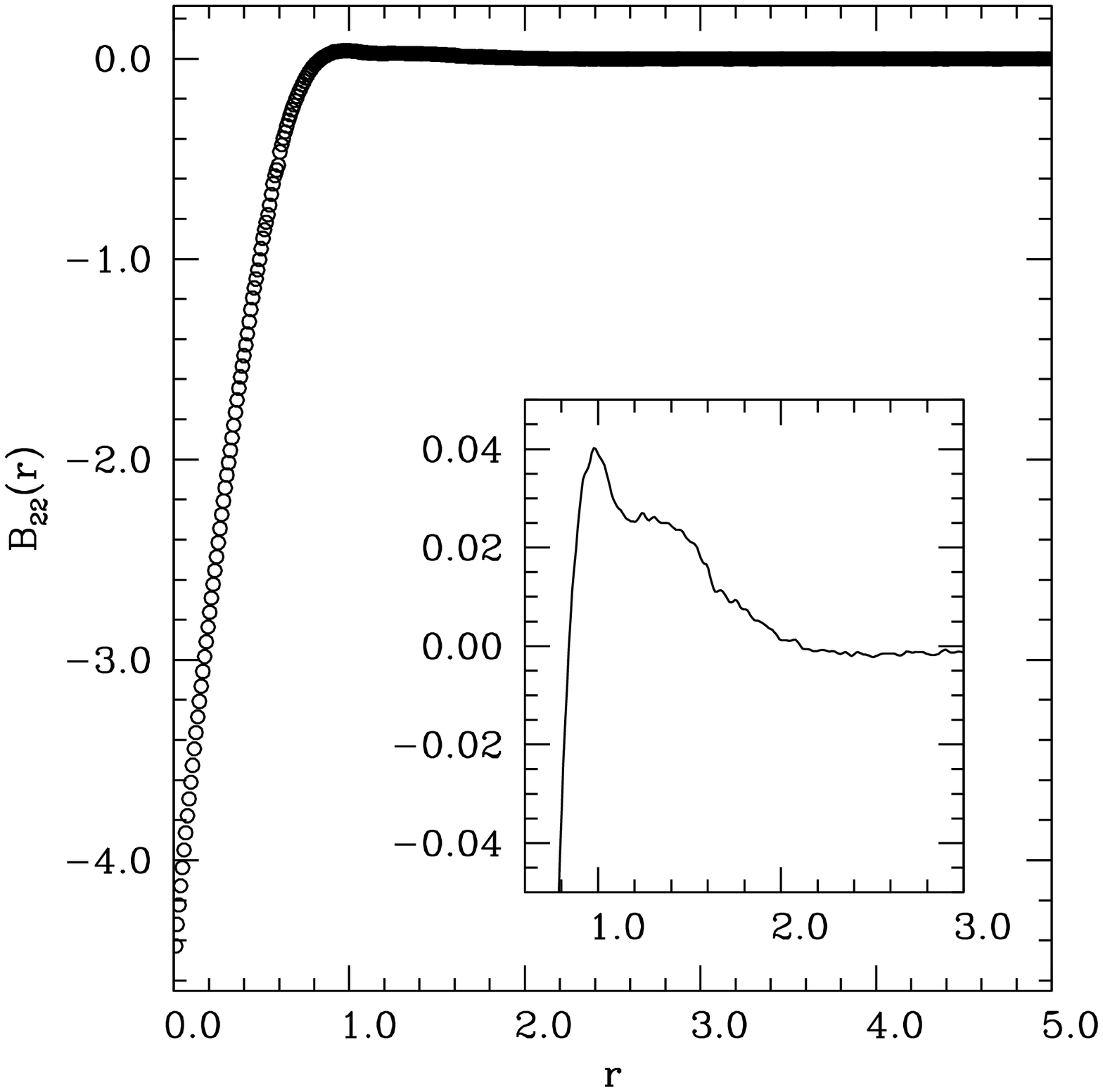}\\
\end{center}
\caption[]{R. Fantoni and G. Pastore
\label{br_ahs-ed}
}
\end{figure}

\begin{figure}[hbtp]
\begin{center}
\includegraphics[width=8cm]{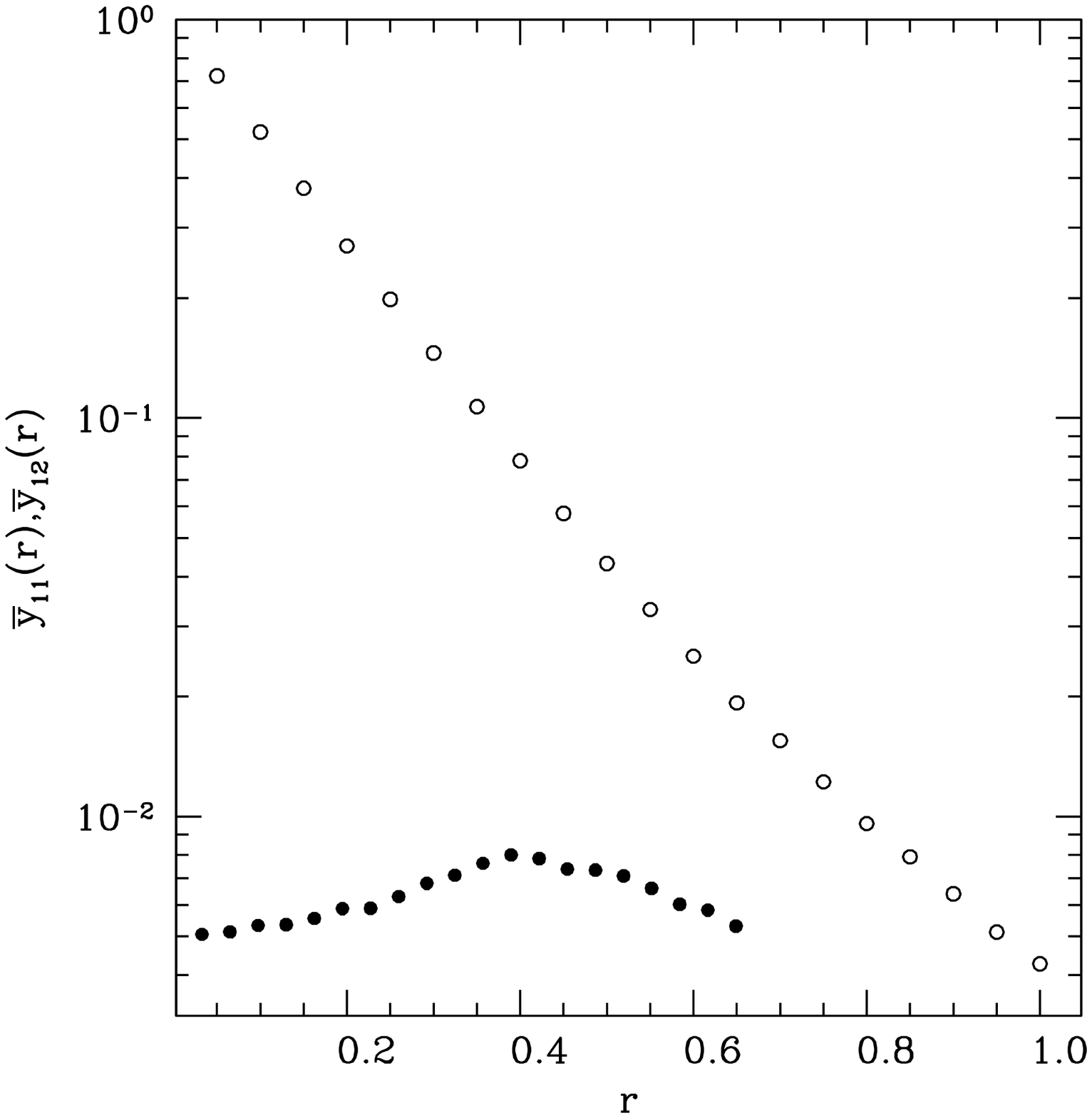}
\end{center}
\caption[]{R. Fantoni and G. Pastore 
\label{bic_nahs-een}
}
\end{figure}

\begin{figure}[hbtp]
\begin{center}
\includegraphics[width=8cm]{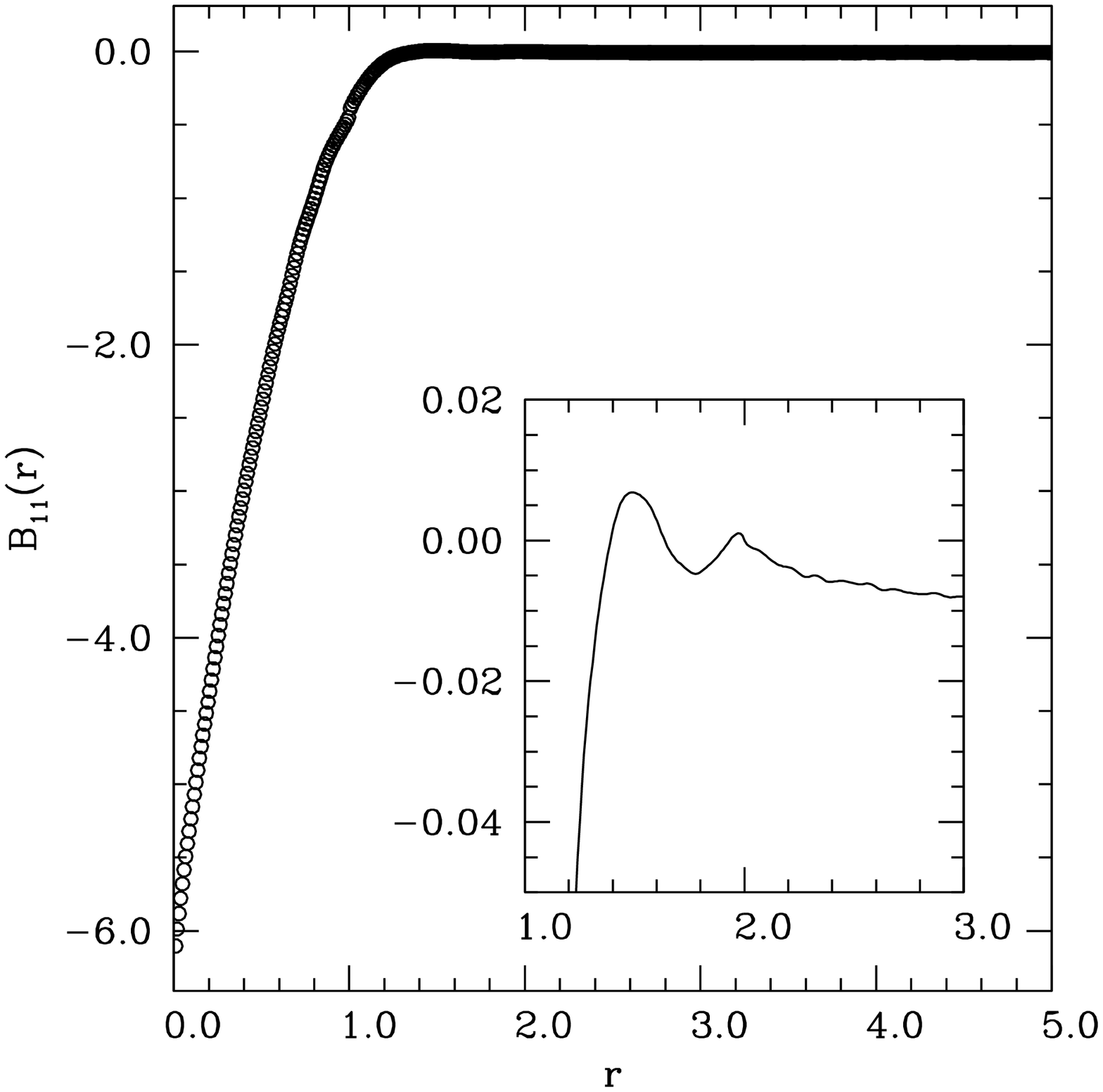}
\includegraphics[width=8cm]{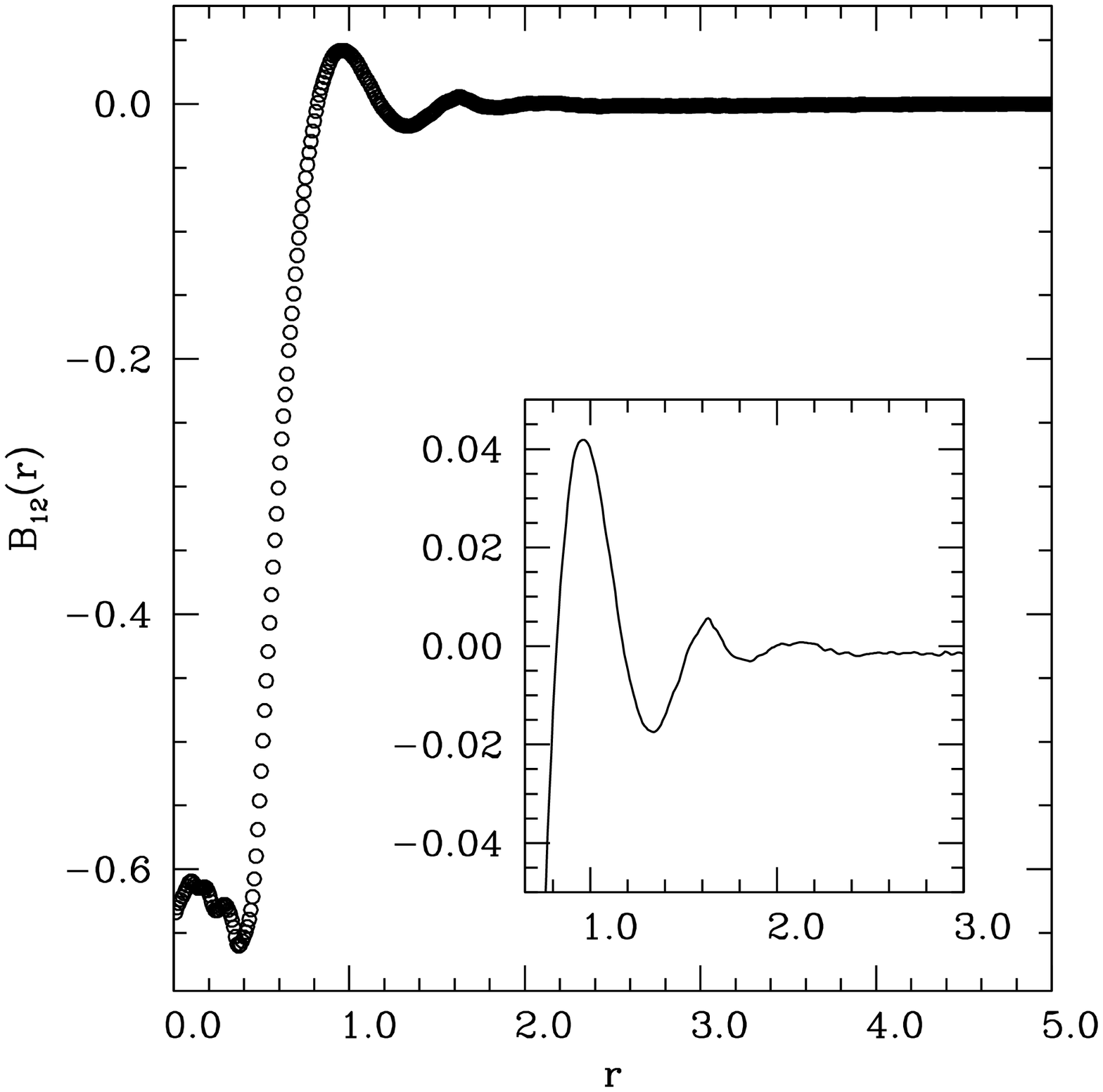}
\end{center}
\caption[]{R. Fantoni and G. Pastore
\label{br_nahs-een}
}
\end{figure}

\begin{figure}[hbtp]
\begin{center}
\includegraphics[width=8cm]{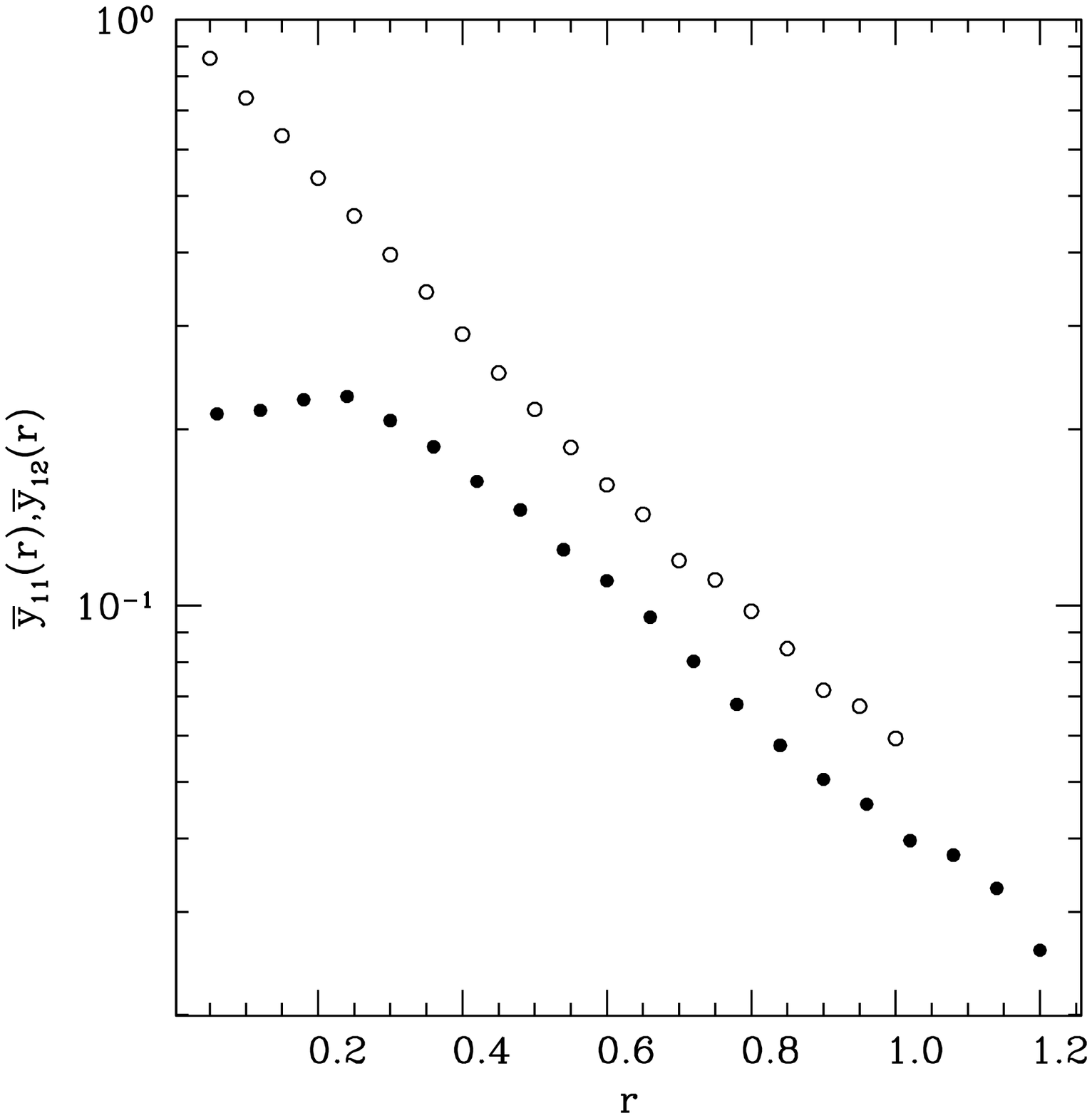}
\end{center}
\caption[]{R. Fantoni and G. Pastore 
\label{bic_nahs-eep}
}
\end{figure}

\begin{figure}[hbtp]
\begin{center}
\includegraphics[width=8cm]{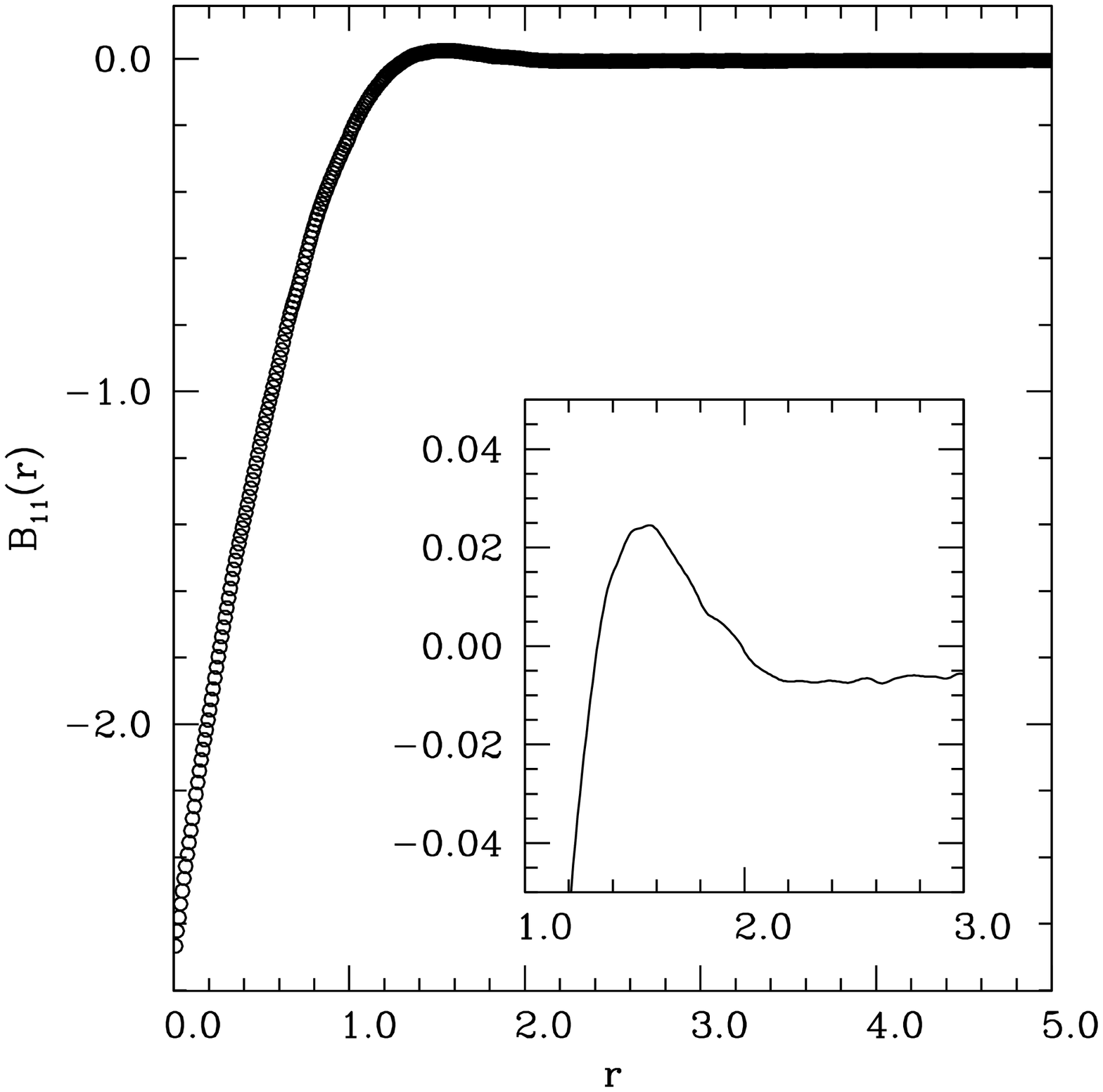}
\includegraphics[width=8cm]{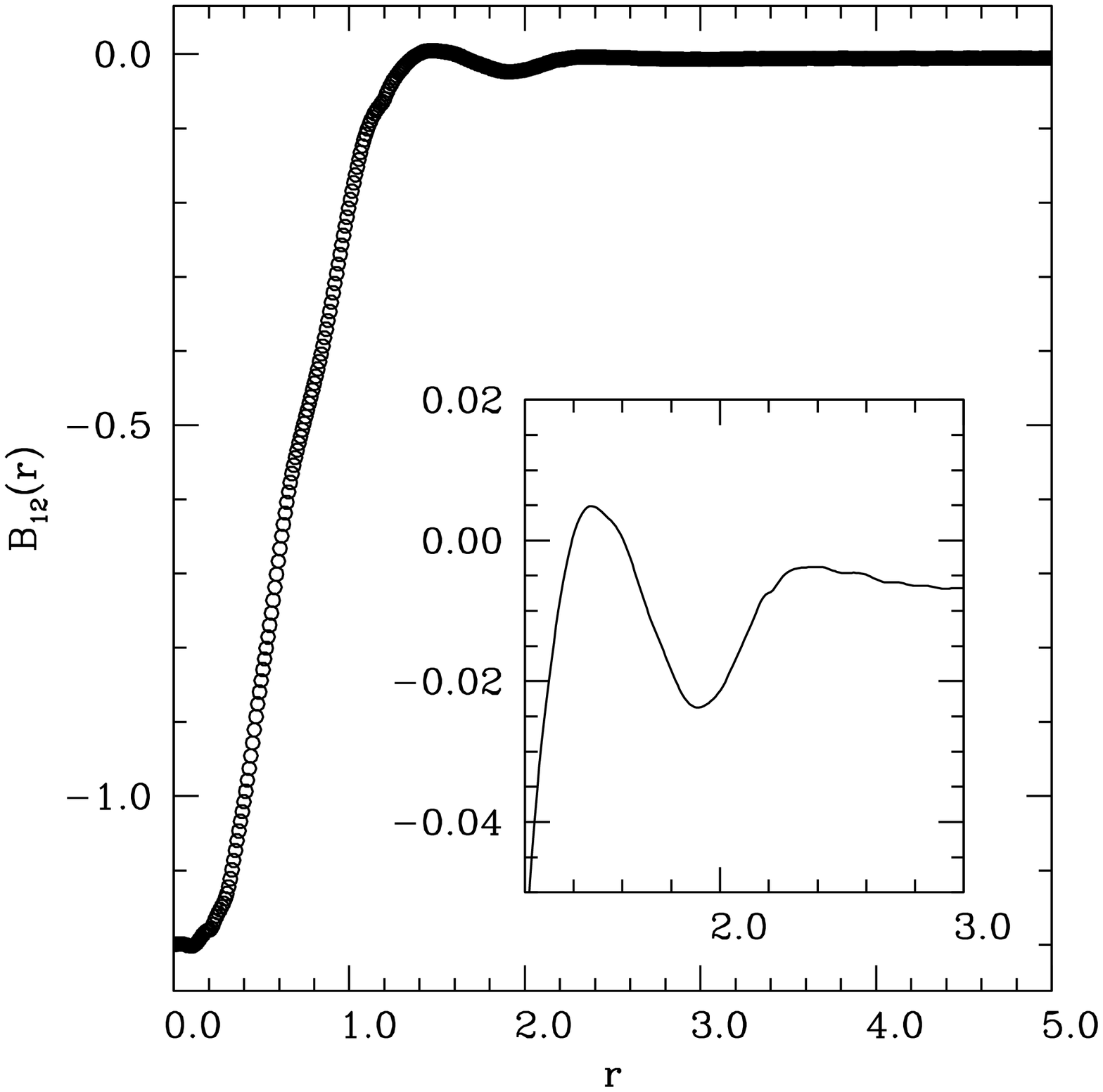}
\end{center}
\caption[]{R. Fantoni and G. Pastore
\label{br_nahs-eep}
}
\end{figure}

\begin{figure}[hbtp]
\begin{center}
\includegraphics[width=8cm]{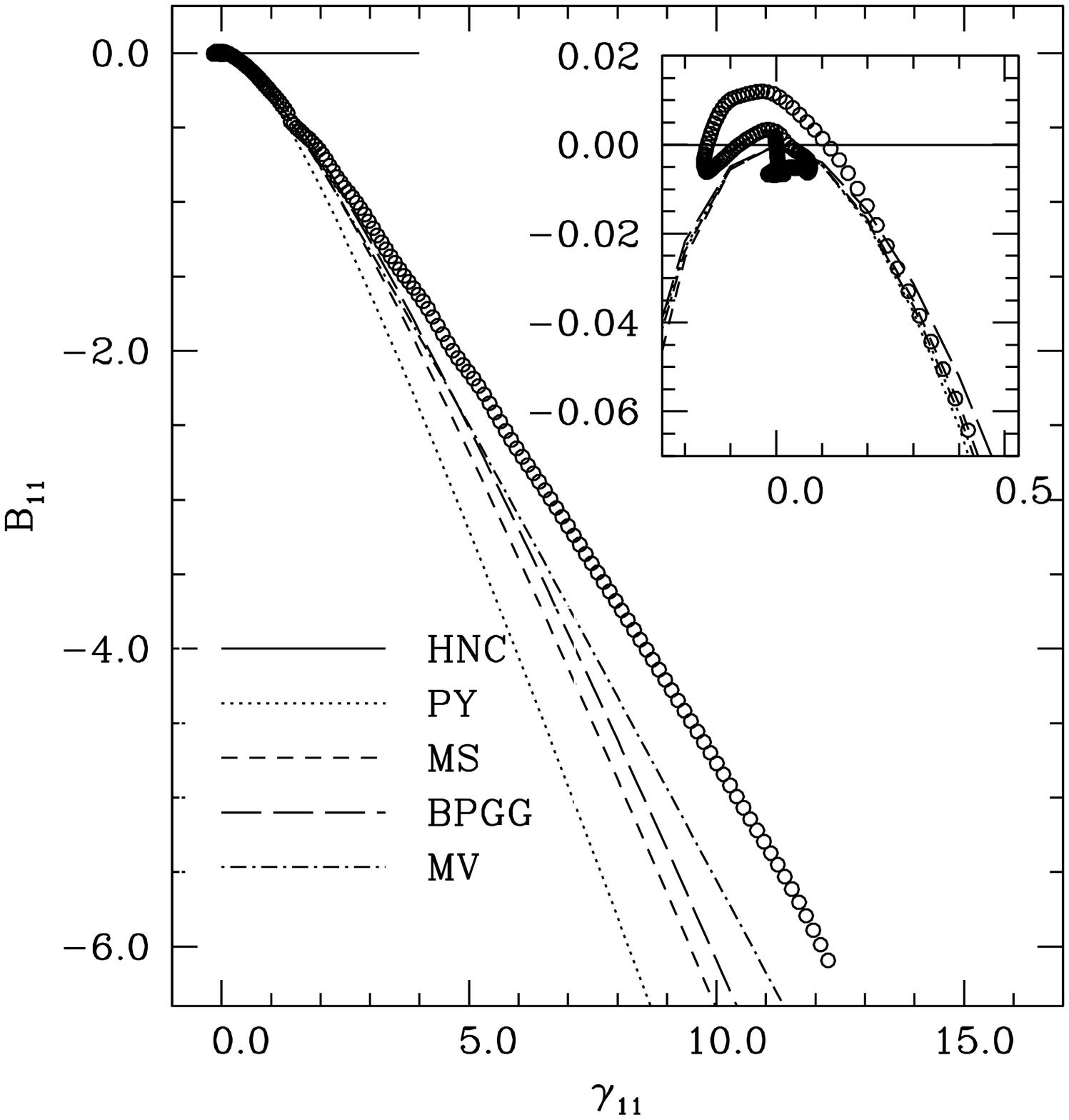}
\includegraphics[width=8cm]{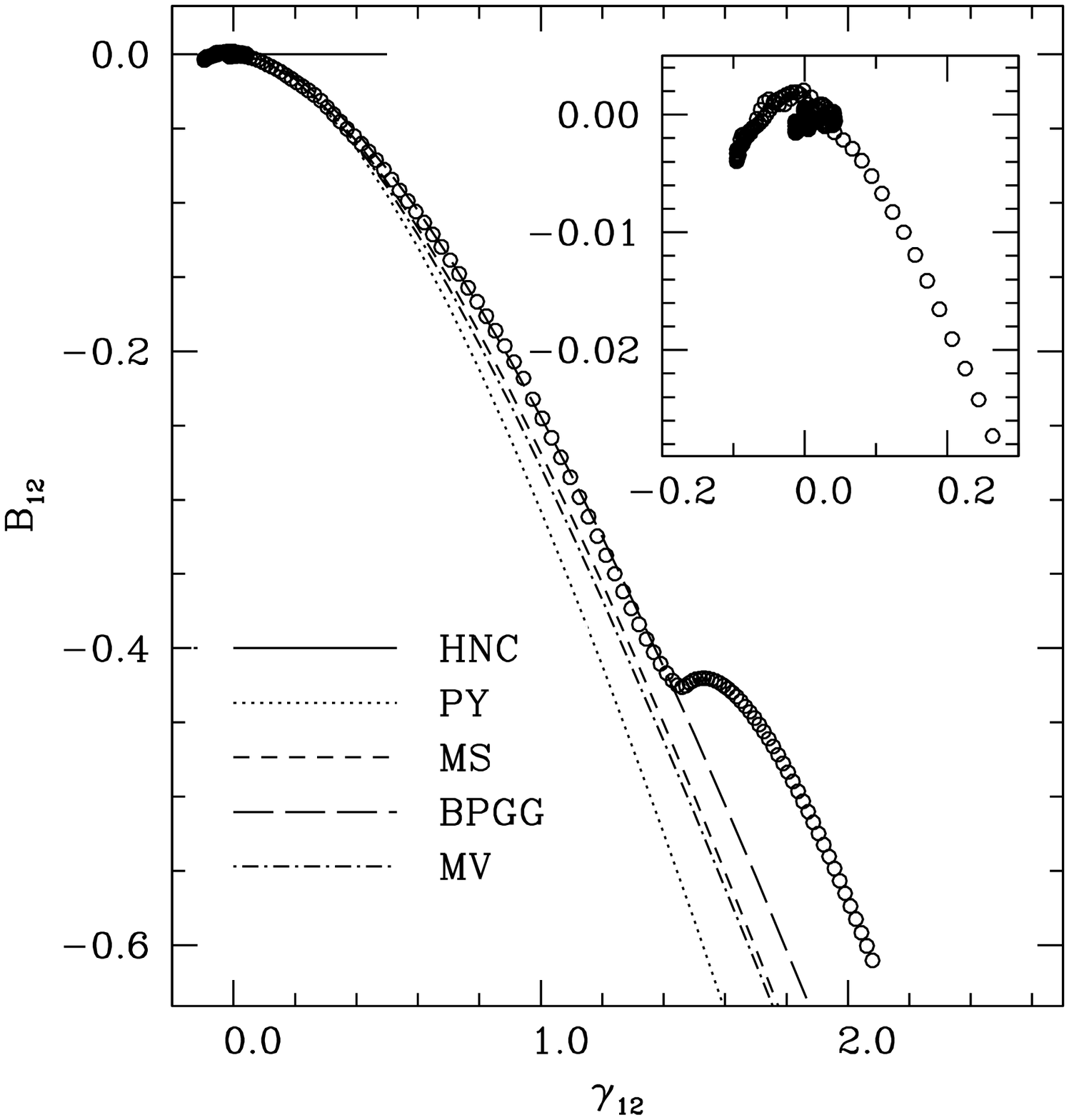}\\
\includegraphics[width=8cm]{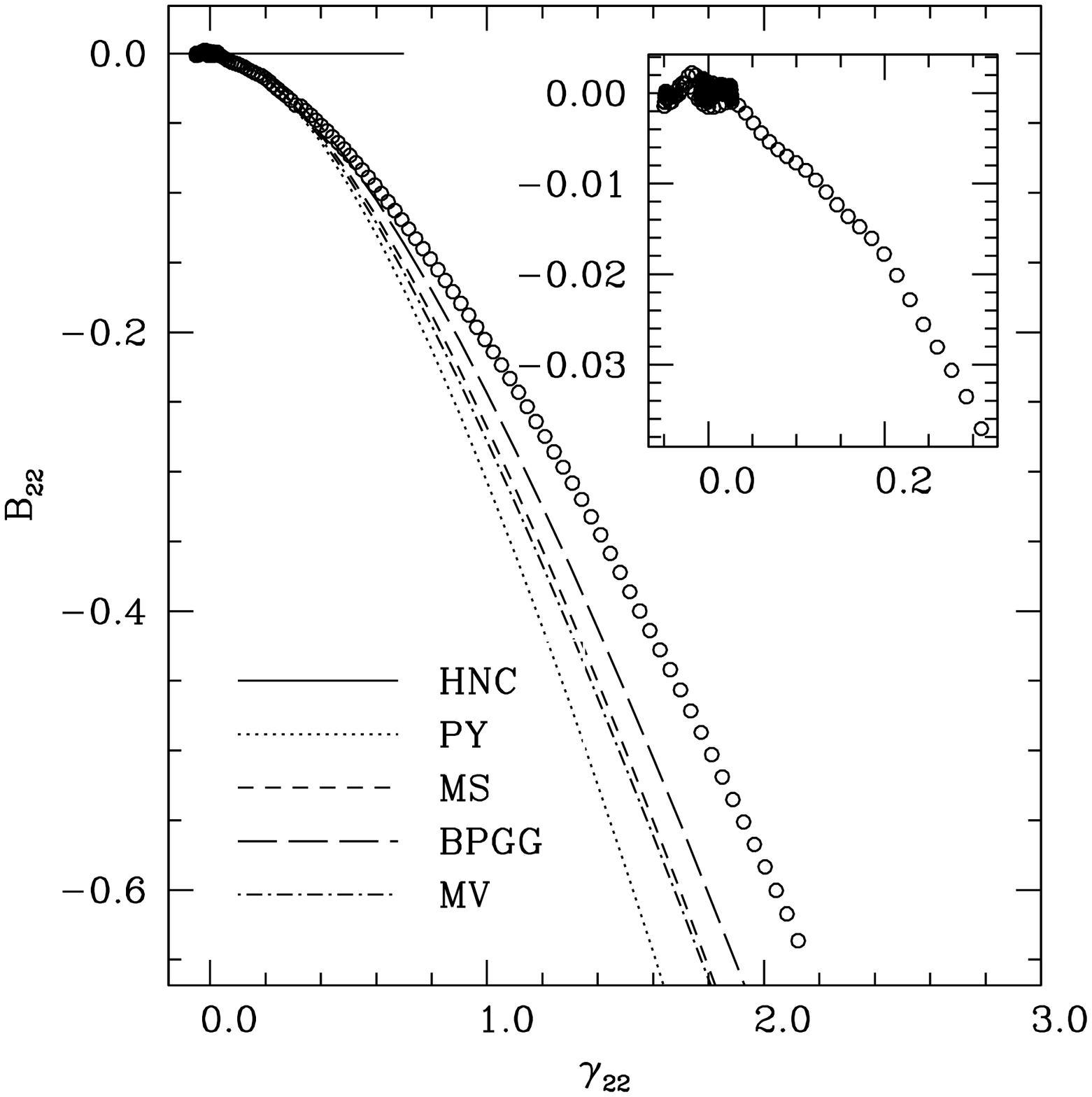}
\end{center}
\caption[]{R. Fantoni and G. Pastore
\label{bvg_nahs-edn}
}
\end{figure}

\begin{figure}[hbtp]
\begin{center}
\includegraphics[width=8cm]{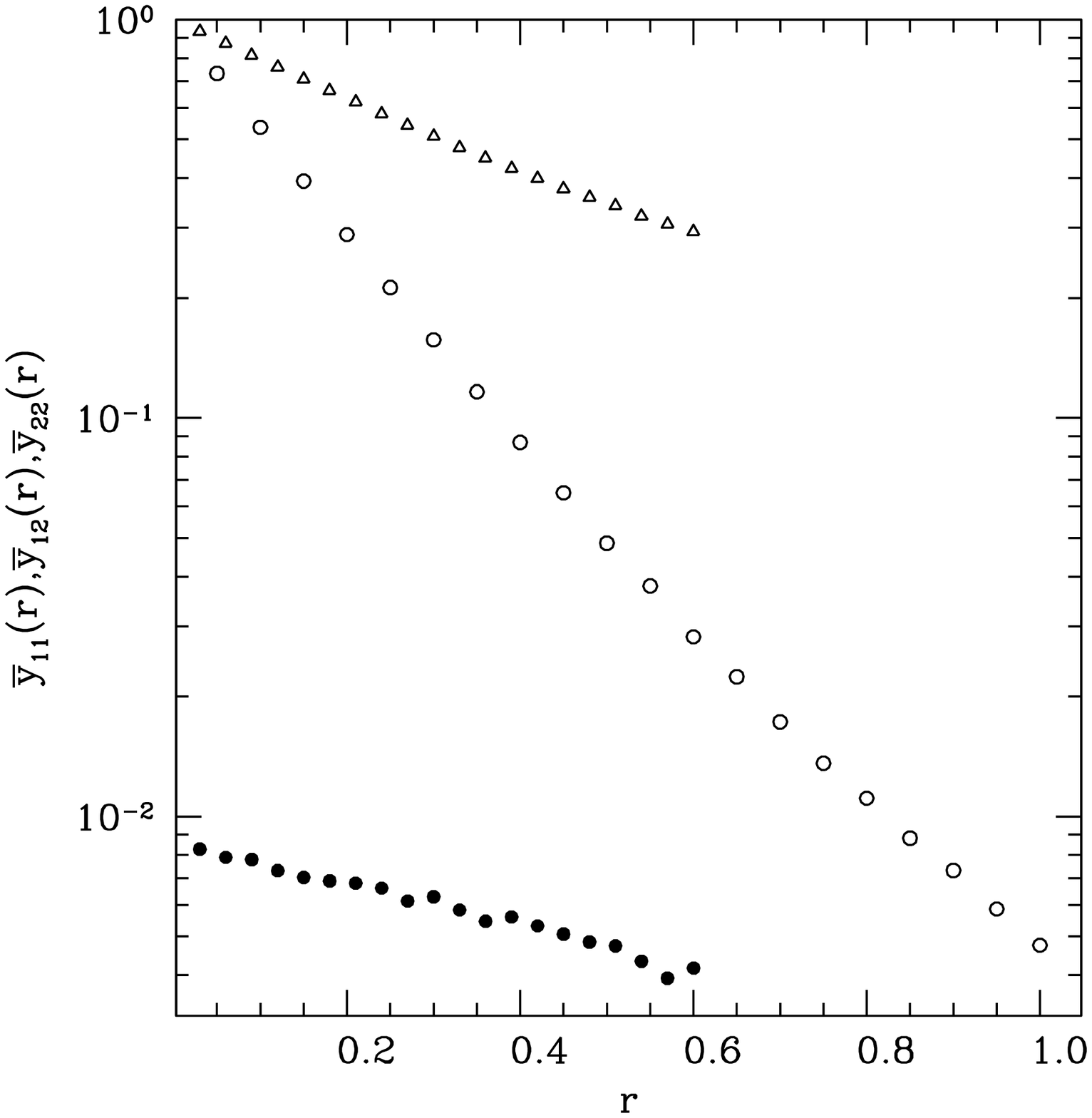}
\end{center}
\caption[]{R. Fantoni and G. Pastore 
\label{bic_nahs-edn}
}
\end{figure}

\begin{figure}[hbtp]
\begin{center}
\includegraphics[width=8cm]{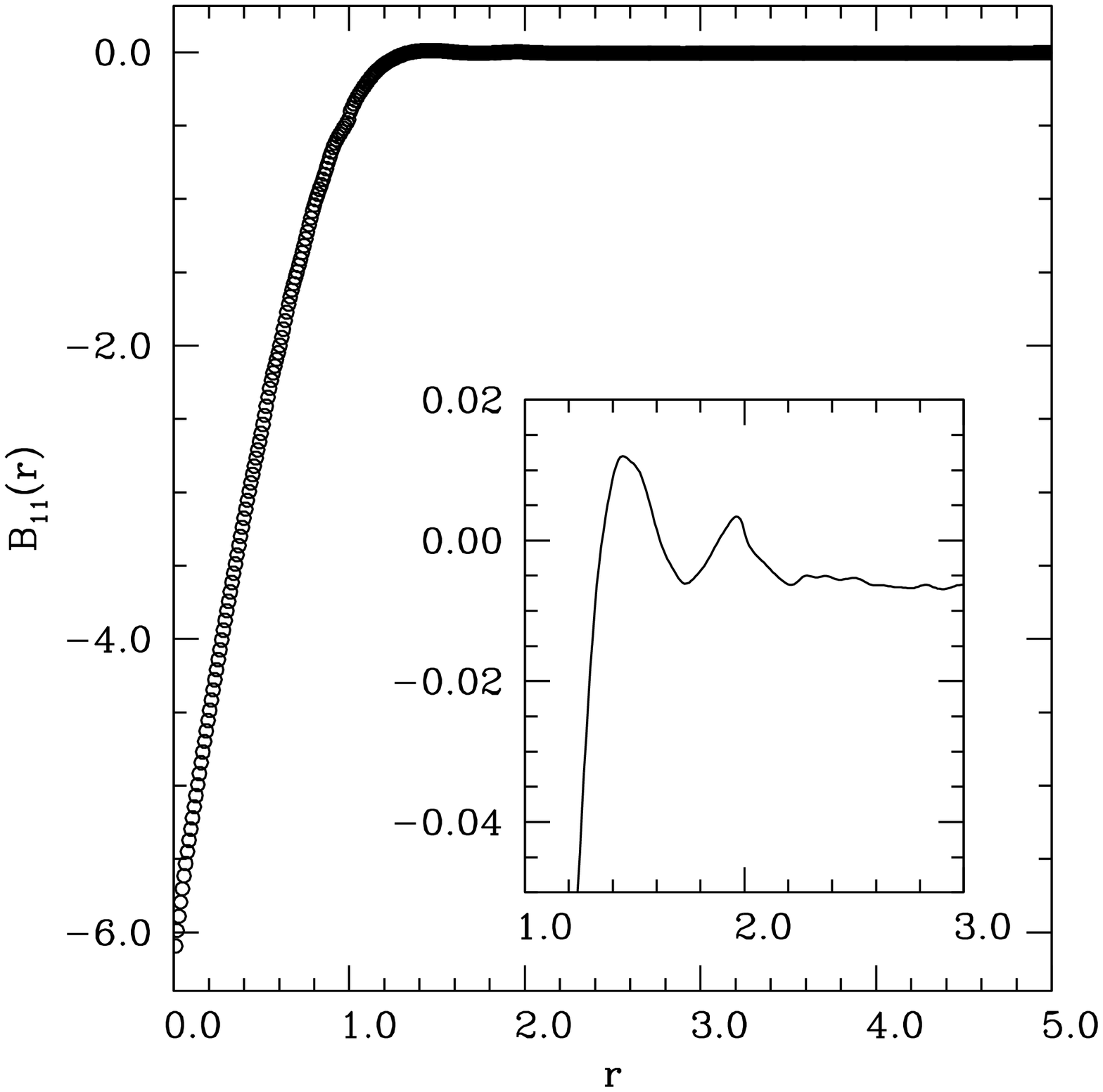}
\includegraphics[width=8cm]{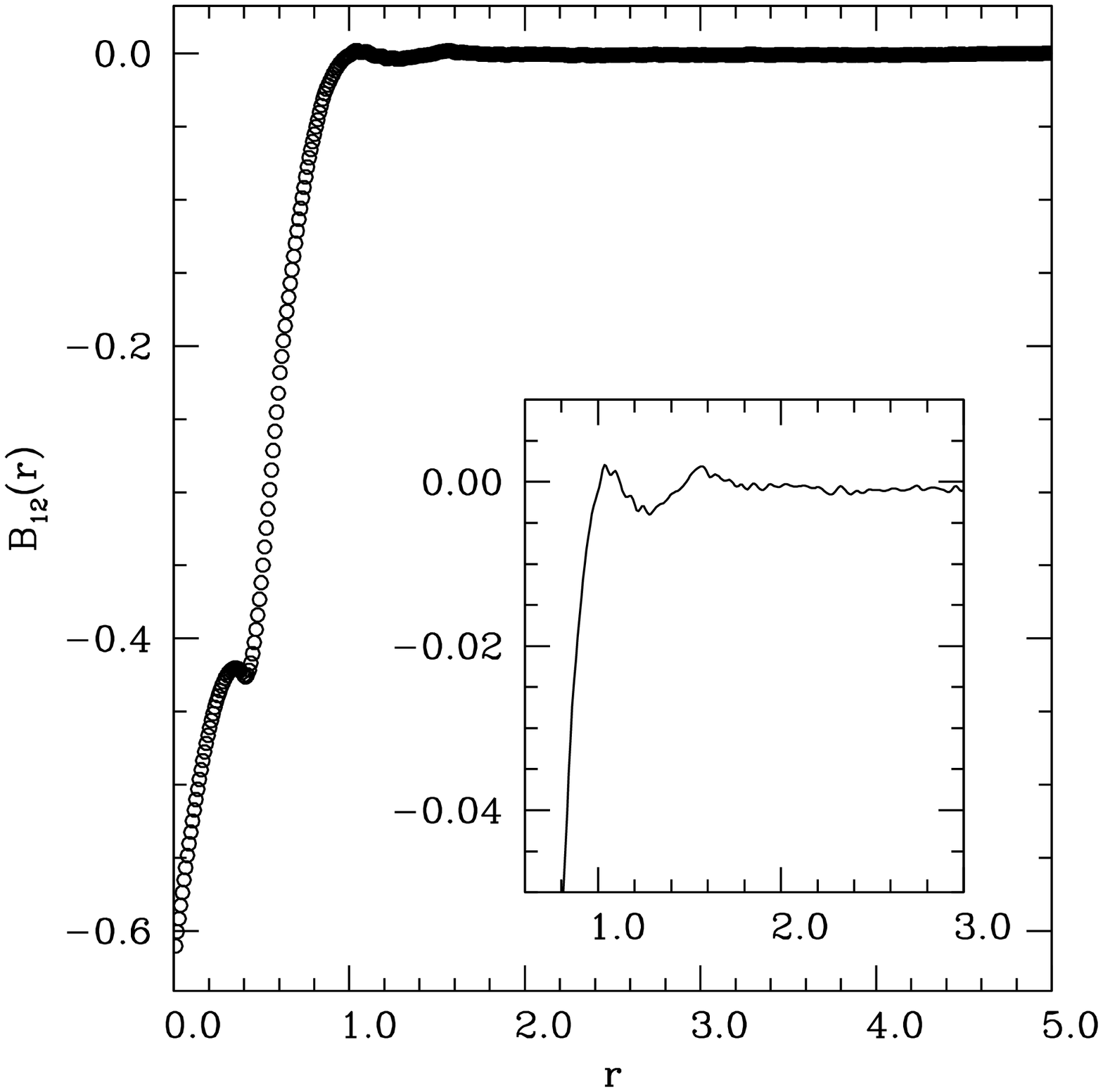}\\
\includegraphics[width=8cm]{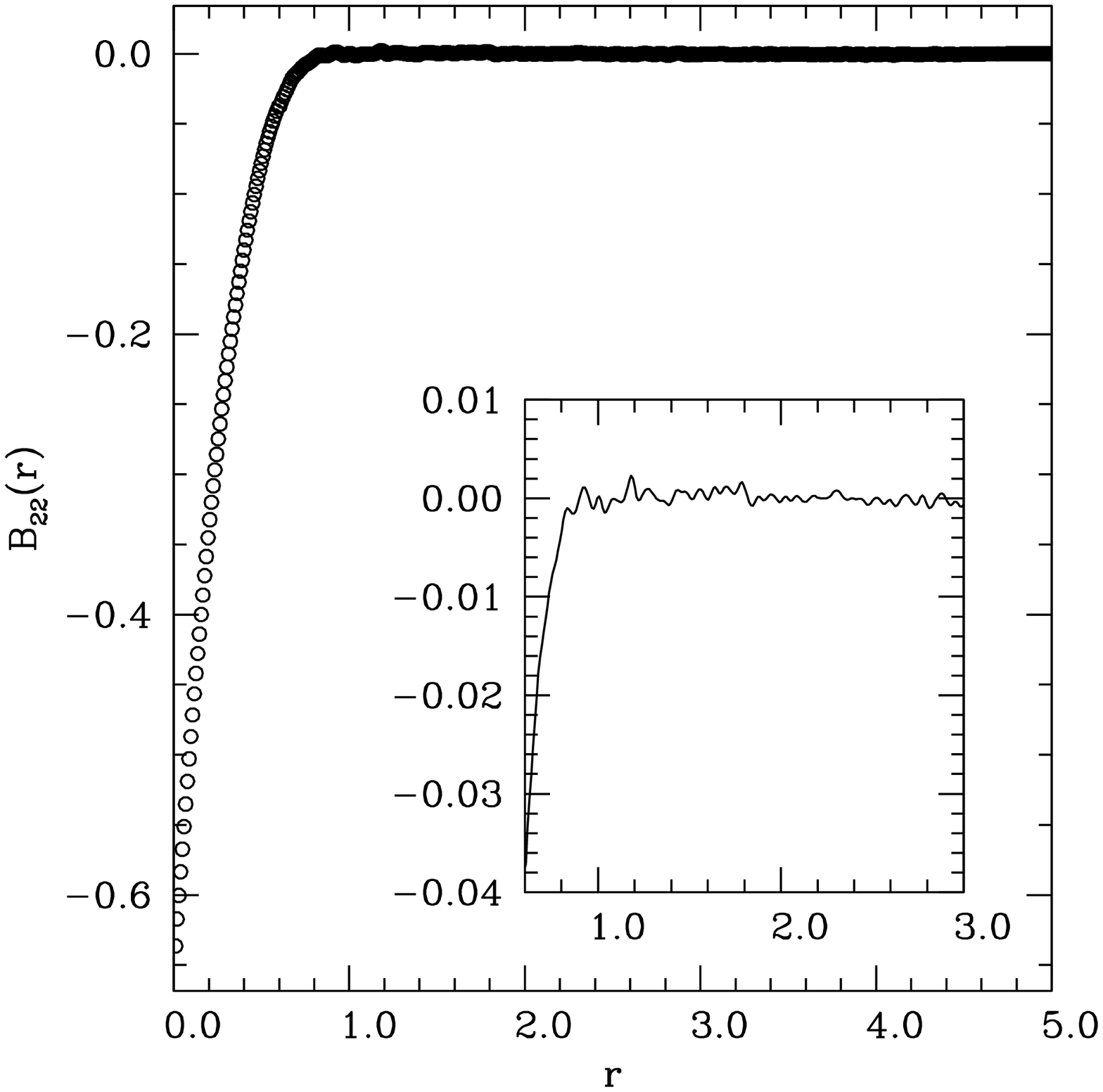}
\end{center}
\caption[]{R. Fantoni and G. Pastore
\label{br_nahs-edn}
}
\end{figure}

\newpage
\centerline{\bf LIST OF FIGURES}
\begin{itemize}
%1
\item[Fig. \ref{boc_ochs}] The first two graphs are Duh-Haymet plots
(dots), outside the hard core region, for the one component HS
system (the lines show the behavior of integral equation closures). 
On the left $\rho=0.650$ on the right $\rho=0.925$.

%2
\item[Fig. \ref{bvg_ahs-ed}] Full Duh-Haymet plots obtained by the
inversion of the Monte Carlo simulation data (dots) compared with some
of the most common integral equation theories (lines) for the
equimolar binary mixture of AHS at two different densities (in the
second and third plot only results at the highest density are shown). 
$R_{11}=1$, $R_{12}=0.8$, and $R_{22}=0.6$. The insets shows the
portion of the bridge function outside the hard cores.

%3
\item[Fig. \ref{bic_ahs-ed}] Cavity functions inside the hard core for
the equimolar binary mixture of AHS (at the same conditions as in
Fig. \ref{bvg_ahs-ed} at the highest density). The plot shows
the behavior of the 
functions defined in (\ref{loc:y11}) and (\ref{loc:y12}) (notice the
logarithmic scale on the ordinates), the triangles denote the 22
function, the open circles the 11 function, and the closed circle the
12 function.
 
%4
\item[Fig. \ref{br_ahs-ed}] Bridge functions $B_{ab}(r)$
for the equimolar binary mixture of AHS (at the same conditions as in
Fig. \ref{bvg_ahs-ed} at the highest density). The insets shows
magnifications of the regions just outside of the hard cores.

%5
\item[Fig. \ref{bvg_nahs-een}] Full
Duh-Haymet plots obtained by the inversion of the Monte Carlo
simulation data (dots) and by some of the most common integral
equation theories (lines) for the equimolar binary mixture of NAHS
with equal like diameters and negative non additivity $\alpha=-0.351$,
at $\rho_1=0.589$. $R_{11}=R_{22}=1$ and $R_{12}=0.649$. The insets
shows the portion of the bridge function outside the hard cores.

%6
\item[Fig. \ref{bic_nahs-een}] Cavity functions for the
equimolar binary mixture of NAHS with equal like diameters and
negative non additivity (at the same conditions as in
Fig. \ref{bvg_nahs-een}). The graph shows the behavior of the 
functions defined in (\ref{loc:y11}) and (\ref{loc:y12}) (notice the
logarithmic scale on the ordinates), the open circle denotes the like
functions and the closed circle the unlike one.

%7
\item[Fig. \ref{br_nahs-een}] Bridge functions $B_{ab}(r)$
for the equimolar binary mixture of NAHS with equal like diameters and
negative non additivity (at the same conditions as in
Fig. \ref{bvg_nahs-een}). The insets shows magnifications of the regions
just outside of the hard cores. 

%8
\item[Fig. \ref{bvg_nahs-eep}] Full
Duh-Haymet plots obtained by the inversion of the Monte Carlo
simulation data (dots) and by some of the most common integral
equation theories (lines) for the equimolar binary mixture of NAHS
with equal like diameters and positive non additivity $\alpha=+0.2$,
at $\rho_1=0.200$. $R_{11}=R_{22}=1$ and $R_{12}=1.2$. The insets
shows the portion of the bridge function outside the hard cores.

%9
\item[Fig. \ref{bic_nahs-eep}] Cavity functions for the
equimolar binary mixture of NAHS with equal like diameters and
positive non additivity (at the same conditions as in
Fig. \ref{bvg_nahs-eep}). The graph shows the behavior of the
functions defined in (\ref{loc:y11}) and (\ref{loc:y12}) (notice the
logarithmic scale on the ordinates), the open circle denotes the like
functions and the closed circle the unlike one.

%10
\item[Fig. \ref{br_nahs-eep}] Bridge functions $B_{ab}(r)$
for the equimolar binary mixture of NAHS with equal like diameters and
positive non additivity (at the same conditions as in
Fig. \ref{bvg_nahs-eep}). The insets shows magnifications of the 
regions just outside of the hard cores. 

%11
\item[Fig. \ref{bvg_nahs-edn}] Full
Duh-Haymet plots obtained by the inversion of the Monte Carlo
simulation data (dots) and by some of the most common integral
equation theories (lines) for the equimolar binary mixture of NAHS
with different like diameters $R_{11}=1$ and $R_{12}=R_{22}=0.6$, at
$\rho_1=0.589$. The insets shows the portion of the bridge function
outside the hard cores.

%12
\item[Fig. \ref{bic_nahs-edn}] Cavity functions for the
equimolar binary mixture of NAHS with different like diameters (at the
same conditions as in Fig. \ref{bvg_nahs-edn}). The graph shows
the behavior of the functions defined in (\ref{loc:y11}) and
(\ref{loc:y12}) (notice the logarithmic scale on the ordinates), the
triangles denote the 22 function, the open circles the 11 function,
and the closed circle the 12 function.

%13
\item[Fig. \ref{br_nahs-edn}] Bridge functions $B_{ab}(r)$
for the equimolar binary mixture of NAHS with different like diameters
(at the same conditions as in Fig. \ref{bvg_nahs-edn}). The insets
shows magnifications of the regions just outside of the hard cores.

\end{itemize} 
\newpage

% fig 1
%

% fig 2

% fig 3
%

% fig 4
%

% fig 5
% 

% fig 6
%

% fig 7
%

% fig 8
%                                                                             

% fig 9
%

% fig 10
%

% fig 11
%

% fig 12
%

% fig 13
%

%%%%%%%%%%%%%%%%%%%%%%%%%%%%%%%%%%%%%%%%%%%%%%%%%%%%%%%%%%%%%%%%%%%%%%%%%%%%%%%
%%%%%%%%%%%%%%%%%%%%%%%%%%%%%%%%%%%%%%%%%%%%%%%%%%%%%%%%%%%%%%%%%%%%%%%%%%%%%%%
%%%%%%%%%%%%%%%%%%%%%%%%%%%%%%%%%%%%%%%%%%%%%%%%%%%%%%%%%%%%%%%%%%%%%%%%%%%%%%%

\end{document}